\documentclass[12pt,a4paper]{article}
\usepackage{amsmath,amssymb,mathtools,amsthm,mathrsfs,multirow,booktabs,setspace,morefloats,epsfig,caption,subfigure}

\usepackage{caption}
\usepackage{natbib}
\bibpunct{[}{]}{;}{n}{,}{,}

\usepackage[utf8]{inputenc}
\usepackage[section]{placeins}
\usepackage{lscape}

\newcounter{list-counter}
\newcounter{algo}[section]
\renewcommand{\thealgo}{\arabic{algo}}
\newenvironment{algo}
{
    \medskip
    \refstepcounter{algo}
    \noindent\textbf{Algorithm~\thealgo}
    \vspace{-7mm}
    \begin{list}{Step~\arabic{list-counter}.~}{\usecounter{list-counter}}{\medskip}
    \setlength{\itemsep}{-1mm}
}
{
    \end{list}
}

\title{\sc Order Restricted Bayesian Analysis of a Simple Step Stress Model 
}
\author{\sc Debashis Samanta$^1$, Debasis Kundu$^2$, Ayon Ganguly$^3$}
\date{}

\begin{document}
\maketitle
\parskip = 10pt

\doublespacing


\begin{center} 
{\sc Abstract}
\end{center}

In this article we consider a simple step stress set up under the cumulative
exposure model  assumption.  At each stress level the lifetime distribution of the experimental units
are assumed to follow the  generalized exponential distribution.  We provide the order restricted
Bayesian inference of the model parameters by considering the fact that the expected lifetime
of the experimental units are larger in lower stress level.  Analysis and the related results are
extended to different censoring schemes also.  The Bayes estimates and the associated credible intervals
of the unknown parameters are constructed using importance sampling technique.  We perform extensive simulation
experiments both for the complete and censored samples to see the performances of the proposed estimators.
We analyze two simulated and one real data sets for illustrative purposes.  An optimal value of the stress changing time is obtained by minimizing 
the total posterior coefficient of variations of the unknown parameters.

\enlargethispage{2 cm}

\noindent$^1$ Department of Statistics, Rabindra Mahavidyalaya, Champadanga, Hooghly, Pin 712401, W.B, India.

\noindent $^2$ Department of Mathematics and Statistics, Indian Institute of Technology Kanpur, Pin 208016, India.  Corresponding
author.

\noindent $^3$ Department of Mathematics, Indian Institute of Technology Guwahati, North Guwahati, Assam 781039, India. 
Allahabad, India.

\section{\sc Introduction}
\paragraph{}
Nowadays, since the products are highly  reliable, it is very difficult to get sufficient
failure time data in a normal condition during a reasonable experimental time.  The accelerated life testing (ALT) procedures are 
proposed to overcome this problem.  The ALT method has been introduced
in a reliability experiment mainly to obtain more failures in a shorter interval of time.    
In an ALT experiment, units are put into a higher stress level than the usual that
ensures early failure of the experimental units.  Interested readers are referred to \citet{N1980} and \citet{B:BN2002} 
for an exposure to different ALT models.
The step stress life test (SSLT) model is a special type of the ALT model in which stress level can be 
changed during the experiment.  In a conventional SSLT, the stress levels are changed at pre-fixed time points. 
Hence, in a conventional SSLT experiment, $n$ experimental units are placed into life testing experiment at an initial stress level $S_1$ 
and then the stress level changes to $S_2, S_3, \ldots, S_{m}$
at prefixed time points $\tau_1, \tau_2, \ldots, \tau_{m-1}$, respectively. If $m=2,$ i.e., in case of only two 
stress levels, the experiment is known as the simple SSLT experiment.   

The data collected from such an SSLT experiment, may then be extrapolated to estimate the underlying distribution of failure
times under normal stress level.
To connect the distributions of lifetime under different stress levels various models have been proposed in the 
literature.  One such model was introduced by \citet{S1966}, and it is known as the cumulative exposure model (CEM).  
The CEM relates the distributions of lifetime under different stress levels by assuming
that the residual life of the experimental units depends only on the cumulative exposure that the units have
experienced, with no memory of how this exposure was accumulated.  Latter this model was extensively studied by \citet{N1980}.
Interested readers are referred to a review article by \citet{B2009} or the recent monograph by \citet{B:KG2017}, 
and the references cited therein.

In this paper we consider a simple step stress model when the lifetime distribution of experimental units follow 
generalized exponential (GE) distribution with the common shape parameter $\alpha$ but different scale parameters 
$\theta_1$ and $\theta_2$ at the two different stress levels.  From now on it is assumed that a GE distribution with the 
shape parameter $\alpha > 0$ and scale parameter $\lambda > 0$, has the following probability density function (PDF)
\begin{equation}
f(t; \alpha, \lambda) = \alpha \lambda (1- e^{-\lambda t})^{\alpha-1} e^{-\lambda t}; \ \ \  t \ge 0,   \label{ge-pdf}
\end{equation}
zero otherwise, and it will be denoted by GE$(\alpha,\lambda)$.  The GE distribution was first considered by \citet{GK1999} 
as an alternative to the well known gamma or Weibull distributions.  It is also an extension of the exponential distribution,
and it also can have increasing or decreasing hazard functions similar to the gamma and Weibull distributions.
The GE distribution has a decreasing density function if the shape parameter is less than one and the density function 
becomes unimodal if the shape parameter is greater than one.  This distribution
has a very good interpretation in case of integer shape parameter.  If the shape parameter is an integer, this distribution
represents the lifetime of a parallel system where each component follows independent exponential distribution.  It is 
observed, see for example \citet{GK2001}, that there are many cases where GE provides a better fit than the gamma
or Weibull distribution.  Interested readers are referred to the article by \citet{NAD2011} for a survey on the GE distribution 
and the recent monograph by \citet{B:AA2015} for the development of the different exponentiated distributions.
It may be mentioned that \citet{HH2009} considered the inference of the parameters of a GE 
distribution for simple SSLT model for Type-I censored data.

In a step stress model the basic assumption is that the expected lifetime of units under higher stress
level is shorter than under the lower stress level. Therefore, this information can be incorporated 
by considering the order restriction on the scale parameters as $\theta_1<\theta_2$.  It seems although for a step-stress 
model, the order restricted inference is a natural choice, not much work has been done along this line mainly due to 
analytical difficulty.  The order restricted inference for an exponential step stress model was first considered by 
\citet{BBK2009} in case of Type-I and Type-II censored data.  It is observed that for exponential model, although the 
maximum likelihood estimators (MLEs) of the unknown parameters can be obtained in explicit forms, the associated exact confidence 
intervals cannot be obtained in explicit form.  Bayesian inference seems to be a reasonable choice in this case.  
\citet{SGKM} developed the order restricted Bayesian inference for exponential simple 
step stress model.  They obtained the Bayes estimates and the associated credible intervals of the unknown parameters 
under the squared error loss function based on importance sampling technique.   The results have been developed for different censoring 
schemes also.

The main aim of this paper is to provide the Bayesian inference on order restricted parameters of a GE distribution for a simple 
SSLT model.  It is assumed that at the two different stress levels the lifetime distributions of the items follow 
GE$(\alpha,\theta_1)$ and GE$(\alpha,\theta_2)$, respectively with $\theta_1 < \theta_2$.  Moreover, it is assumed that it 
satisfies the CEM assumptions.  We consider the Bayesian inference
on the unknown parameters under a fairly flexible prior assumptions (the details of the priors will be provided in the 
next section).  First
we consider the complete sample, and provide the Bayes estimates and the associated credible intervals based on importance 
sampling technique.  The necessary theoretical results for the convergence of the corresponding importance sampling procedure are also
provided.  The results are extended for different other censoring schemes, namely for Type I censoring, Type II 
censoring, Type I hybrid censoring scheme (HCS), introduced by \citet{E1954}, and for Type II hybrid censoring scheme, 
introduce by \citet{CCBK2003}, also.  Extensive Monte Carlo simulations are performed for complete and censored samples to see 
the performance of the proposed method, and they are quite satisfactory.  Two simulated and one real data sets 
have been analyzed for illustrative purposes.

Finally we consider the `optimal' simple SSLT model under the same assumptions.  Similar to the idea proposed by \citet{ZM2005}, we propose to choose the `optimal' value of $\tau_1$, so that the sum of the posterior coefficient of variations of 
$\alpha$, $\theta_1$ and $\theta_2$ is minimum.  Since the posterior coefficient of variations of the unknown parameters cannot be obtained
in explicit forms, we use Lindley's approximation for the posterior coefficient of variations, and provide a methodology to choose the 
`optimum' $\tau_1$.  A small table with the `optimal' values of $\tau_1$ is provided for different sample sizes and for different
parameter values.

The rest of the paper is organized as follows.  In Section 2, we provide the model and the necessary prior assumptions.  The Bayesian
inference of the unknown parameters for complete sample is provided in Section 3, and for different censoring schemes the results
are provided in Section 4.  Simulation and data analysis results are reported in Section 5.  
In Section 6 we consider the optimality of the simple SSLT model, and finally we conclude the paper in Section 7.

\section{\sc Model Assumption and Prior Information}

Consider the simple step-stress model with two stress levels $S_{1}$ and $S_{2}$. Suppose $n$ items are put
into an experiment under the stress level $S_1$ and the stress level is changed to $S_2$ at a pre-fixed time
$\tau_1$. The failure times, denoted by $t_{1:n}<t_{2:n}<t_{3:n}<\ldots<t_{n:n}$, of the unit placed on the
test are recorded chronologically. It is assumed that the lifetimes have a generalized exponential (GE)
distribution under both the stress levels, with the common shape parameter $\alpha$ and different scale
parameters, say $\theta_1$ and $\theta_2$ under stress level $S_1$ and $S_2$, respectively.  
It is further assumed that the lifetime satisfies CEM assumptions.  Hence, the cumulative distribution function (CDF) of 
the lifetime is given by
\begin{align} \label{eq:CDF}
        F(t) = \left\{ \begin{array}{ll}
                        (1-e^{-\theta_1 t})^\alpha & \text{ if } 0 < t \le \tau_1 \\
                        (1-e^{-\theta_2(t+\frac{\theta_1}{\theta_2}\tau_1-\tau_1)})^\alpha & \text{ if }
                        \tau_1 < t < \infty,
                    \end{array}
                \right.
\end{align}
and corresponding PDF is given by
\begin{align}\label{eq:PDF}
    f(t) = \left\{
                \begin{array}{ll}
                    \alpha\theta_1(1-e^{-\theta_1 t})^{\alpha-1}e^{-\theta_1t} & \text{ if } 0 < t \le \tau_1\\
                    \alpha\theta_2(1-e^{-\theta_2(t+\frac{\theta_1}{\theta_2}\tau_1-\tau_1)})^{\alpha-1}
                    e^{-\theta_2(t+\frac{\theta_1}{\theta_2}\tau_1-\tau_1)} & \text{ if } \tau_1 < t < \infty.
                \end{array}
            \right.
\end{align}
The purpose of an ALT procedure is to increase the stress level which ensures the early failure of
the experimental units. Hence, it is reasonable to assume that the mean lifetime at the stress level $S_1$ is larger
than that at the stress level $S_2$, i.e.,
\begin{equation}
\frac{1}{\theta_2}[\psi(\alpha+1)-\psi(1)] < \frac{1}{\theta_1}[\psi(\alpha+1)-\psi(1)],  \label{exp-life}  
\end{equation}
 where $\displaystyle \psi(\cdot)$ is the digamma function.  From (\ref{exp-life}), it follows that $\theta_1<\theta_2$.  
We use this information in our prior assumption as follows.  Let us assume 
$\theta_1 = \beta\theta_2$, where $0 < \beta <1$. Suppose the prior belief of the experimenter is measured by
the function $\pi(\alpha,\theta_2,\beta)$, which is  given by
$$\pi(\alpha,\theta_2,\beta) = \pi_1(\alpha)\pi_2(\theta_2)\pi_3(\beta).$$
It is assumed that $\alpha\sim Gamma(a_0,b_0)$, $\theta_2\sim Gamma(a_1,b_1)$,
$\beta\sim Beta(a_2,b_2)$ and they are independently distributed.  
The joint prior distribution of $(\alpha, \theta_2, \beta)$, is given by
\begin{align}
\pi(\alpha,\theta_2,\beta) \propto \beta^{a_2-1}(1-\beta)^{b_2-1}
e^{-a_0\alpha}\alpha^{b_0-1}e^{-a_1\theta_2}\theta_2^{b_1-1}.     \label{joint-prior}
\end{align}

\section{\sc Posterior Analysis and Bayesian Inference}

Based on the joint prior distribution (\ref{joint-prior}), and under the CEM assumptions, the joint posterior distribution 
of $\alpha$, $\theta_2$ and $\beta$ is
given by
\begin{align}
l(\beta,\theta_2,\alpha\vert Data) \propto &
\quad\beta^{n_1+a_2-1}(1-\beta)^{b_2-1}\theta_2^{n+b_1-1}e^{-A_1(\beta)\theta_2}\alpha^{n+b_0-1}e^{-A_2(\beta,\,\theta_2)\alpha}\nonumber\\
{} & \quad\times\prod_{i=1}^{n_1}(1-e^{-\beta\theta_2 t_{i:n}})^{-1}\prod_{i=n_1+1}^{n}(1-e^{-\theta_2(t_i-\tau_1+\tau_1\beta)})^{-1},\label{eq:post1}
\end{align}
where $n_1$ denotes the number of failures till $\tau_1$, and 
\begin{align*}
{} & A_1(\beta) = a_1+\beta\sum_{i=1}^{n_1}t_{i:n} + \sum_{i=n_1+1}^n(t_{i:n}-\tau_1+\tau_1\beta),\\
{} & A_2(\beta,\,\theta_2) = a_0-\sum_{i=1}^{n_1}log(1-e^{-\beta\theta_2
t_{i:n}})-\sum_{i=n_1+1}^{n}log(1-e^{-\theta_2(t_{i:n}-\tau_1+\tau_1\beta)}).
\end{align*}
Therefore, the Bayes estimate of some parametric function of $(\beta,\,\theta_2,\,\alpha)$, say
$g(\beta,\,\theta_2,\,\alpha)$, under the squared error loss function is
\begin{align}\label{eq:est1}
\hat{g}_{B}(\beta,\theta_2,\alpha) & = E_{\beta,\theta_2,\alpha\vert Data}\left(g(\beta, \theta_2, \alpha)\right) \nonumber \\
{} &
=\frac{\displaystyle\int_{0}^{1}\int_{0}^{\infty}\int_{0}^{\infty}g(\beta, \theta_2, \alpha)l(\beta,\theta_2,\alpha\vert
Data)d\alpha d\theta_2
d\beta}{\displaystyle\int_{0}^{1}\int_{0}^{\infty}\int_{0}^{\infty}l(\beta,\theta_2,\alpha\vert Data)d\alpha
d\theta_2 d\beta},
\end{align}
provided the expectation exists.  In general \eqref{eq:est1} cannot be obtained in explicit form.  One can use
approximation procedure like Lindley's approximation or Tierney and Kadane's Method.  However, the associated
credible interval cannot be constructed using these techniques.  Hence, we propose to use  importance sampling technique to
compute the Bayes estimates and the associated credible intervals of the unknown parameters.  Note that posterior density of
$(\beta,\,\theta_2,\,\alpha)$ can be written as
\begin{align}
{} & l(\beta,\,\theta_2,\,\alpha\,\vert\,Data) \propto  h(\beta,\,\theta_2,\,\alpha) l_1(\beta)
l_2(\theta_2\,\vert\,\beta) l_3(\alpha\,\vert\,\theta_2,\,\beta),\label{eq:post2}
\end{align}
\begin{align}
\begin{array}{llll}
\shortintertext{where}
{} & h(\beta,\theta_2,\alpha) &= &\beta^{n_1+a_2-1}(1-\beta)^{b_2-1}\left[A_1(\beta)\right]^{-(n+b_1)} \left[A_2(\beta,\,\theta_2)\right]^{-(n+b_0)} \nonumber \\ 
{} &{}&{}& \prod_{i=1}^{n_1}(1-e^{-\beta\theta_2 t_i})^{-1}
\prod_{i=n_1+1}^{n}(1-e^{-\theta_2(t_i-\tau_1+\tau_1\beta)})^{-1}, \nonumber\\
{} & l_1(\beta)& = & 1 \quad\text{ for } 0<\beta<1,\nonumber \\
{} & l_2(\theta_2\vert\beta) & = & 
\frac{\left[A_1(\beta)\right]^{n+b_1}}{\Gamma(n+b_1)}\,\theta_2^{n+b_1-1}\,e^{-A_1(\beta)\,\theta_2} \quad\text{ for }
\theta_2>0, \nonumber \\
{} & l_3(\alpha\vert\theta_2,\beta) & = &
\frac{\left[A_2(\beta,\,\theta_2)\right]^{n+b_0}}{\Gamma(n+b_0)}\,\alpha^{n+b_0-1}\,e^{-A_2(\beta,\,\theta_2)\,\alpha}\quad
\text{ for } \alpha>0.\nonumber
\end{array}
\end{align}
Using equation \eqref{eq:post2}, following algorithm can be executed to compute the Bayes estimate 
and the associated  credible interval of some
parametric function $g(\beta,\,\theta_2,\,\alpha)$ of $\beta, \theta_2$ and $\alpha$, as given in \eqref{eq:est1}. 

\begin{algo}
\item Generate $\beta_1$ from Uniform$(0,1)$, $\theta_{21}$ from
Gamma$\displaystyle\left(n+b_1,\,A_1(\beta_1)\right)$, and $\alpha_1$ from
Gamma$\displaystyle\left(n+b_0,\,A_2(\beta_1,\,\theta_{21})\right)$ distribution.

\item Repeat Step 1, $N$ times to obtain
$(\beta_1,\theta_{21},\alpha_1),\,\ldots,\,(\beta_N,\theta_{2N},\alpha_N)$, where $\beta_i$, $\theta_{2i}$ and $\alpha_i$
is the generation of $\beta$, $\theta_{2}$ and $\alpha$ at $i$-th ($i=1,\ldots,N$) replication respectively. 

\item Calculate $g_i = g(\beta_i,\theta_{2i},\alpha_i)$ and $w_i = \displaystyle
\frac{h(\beta_i,\theta_{2i},\alpha_i)}{\sum_{j=1}^{N}h(\beta_j,\theta_{2j},\alpha_j)}$ for $i = 1, \ldots, N$.

\item The approximate value of \eqref{eq:est1} can be obtained as $\sum_{i=1}^{N} w_i g_i$.

\item Rearrange $(g_1,w_1),\,(g_2,w_2),\,\ldots,\,(g_N,w_N)$ as
$(g_{(1)},w_{(1)}),\,(g_{(2)},w_{(2)}),\,\ldots,\,(g_{(N)},w_{(N)})$ where $g_{(1)}\leq g_{(2)}\leq\ldots\leq
g_{(N)}$. Note that $w_{(i)}$'s are not ordered, they are just associated with $g_{(i)}$'s.

\item A $100(1-\gamma)\%$ credible interval for $g(\beta,\,\theta_2,\,\alpha)$ can be obtain as
$\left(g_{j_1},\,g_{j_2}\right)$, where $j_1$ and $j_2$ satisfy
\begin{align}\label{eq:hpdcri}
j_1,\,j_2\in\left\{1,\,2,\,\ldots,\,N\right\}, \quad j_1<j_2, \quad\sum_{i=j_1}^{j_2}w_{(i)}\leq 1-\gamma<\sum_{i=j_1}^{j_2+1}w_{(i)}.
\end{align}
The $100(1-\gamma)\%$ HPD credible interval (CRI) of $g(\beta,\,\theta_2,\,\alpha)$ becomes
$\left(g_{(j_1^*)},\,g_{(j_2^*)}\right)$, where $1\leq j_1^*<j_2^*\leq N$ satisfy
\begin{align*}
\sum_{i=j_1^*}^{j_2^*}w_{(i)}\leq 1-\gamma<\sum_{i=j_1^*}^{j_2^*+1}w_{(i)}, \quad\text{and}\quad
g_{(j_2^*)}-g_{(j_1^*)}\leq g_{(j_2)}-g_{(j_1)},
\end{align*}
for all $j_1$ and $j_2$ satisfying \eqref{eq:hpdcri}.
\end{algo}

\section{\sc Different Censoring Schemes and Posterior Analysis}
Due to the experimental time and budget restrictions, the experimenter often terminates the experiment before the last unit 
fails.  This is known as censoring in the statistical terminology.  In this section we discuss different censoring schemes and 
associated posterior analysis based on the same prior and model assumptions.  Consider the following general notations for 
different censoring schemes. $n_1^* = $ number of failure before $\tau_1 ;$ $n_2^* = $ number of failure between $\tau_1$ and 
$\tau^* ;$
$\tau^* = $ termination time of the experiment; $n^* = $ total number of failure before $\tau^*.$
\subsection{\sc Type-I Censoring}
In Type-I censoring scheme we stop the experiment at a prefix time, say $\tau_2$ and the number of observations failed under stress level $S_1$
and $S_2$ are $n_1$ and $n_2$ respectively. In this case observed data are one of the forms \newline
$(a)$ \quad $\{\tau_1<t_{1:n}<...<t_{n_2:n}<\tau_2\}$, \newline 
$(b)$ \quad $\{t_{1:n}<t_{2:n}<...<t_{n_1:n}<\tau_1<t_{n_1+1:n}<...<t_{n_1+n_2:n}<\tau_2 \}$, \newline 
$(c)$ \quad $\{t_{1:n}<t_{2:n}<...<t_{n_1:n}<\tau_1<\tau_2$\}. \newline 
Under Type-I censoring scheme posterior distribution can be written as
\begin{align}
{} & l(\beta,\,\theta_2,\,\alpha\,\vert\,Data) \propto h_1(\beta,\,\theta_2,\,\alpha) l_1(\beta)
l_2(\theta_2\,\vert\,\beta) l_3(\alpha\,\vert\,\theta_2,\,\beta),\label{eq:post3}
\end{align}
\begin{align}
\begin{array}{llll}
\shortintertext{where}
{} & h_1(\beta,\theta_2,\alpha) & = & \beta^{n_1^*+a_2-1}(1-\beta)^{b_2-1}\left[A_1(\beta)\right]^{-(n^*+b_1)} \left[A_2(\beta,\,\theta_2)\right]^{-(n^*+b_0)}
\left[A_3(\beta,\,\theta_2, \alpha)\right]^{(n-n^*)} \nonumber \\
{} & {} & {} & \prod_{i=1}^{n_1^*}(1-e^{-\beta\theta_2 t_i})^{-1}
\prod_{i=n_1^*+1}^{n^*}(1-e^{-\theta_2(t_i-\tau_1+\tau_1\beta)})^{-1}, \nonumber\\
{} & l_1(\beta) & = & 1 \quad\text{ for } 0<\beta<1,\nonumber \\
{} & l_2(\theta_2\vert\beta) & = & 
\frac{\left[A_1(\beta)\right]^{n^*+b_1}}{\Gamma(n^*+b_1)}\,\theta_2^{n^*+b_1-1}\,e^{-A_1(\beta)\,\theta_2} \quad\text{ for }
\theta_2>0, \nonumber \\
{} & l_3(\alpha\vert\theta_2,\beta) & = &
\frac{\left[A_2(\beta,\,\theta_2)\right]^{n^*+b_0}}{\Gamma(n^*+b_0)}\,\alpha^{n^*+b_0-1}\,e^{-A_2(\beta,\,\theta_2)\,\alpha}\quad
\text{ for } \alpha>0,\nonumber \\	
{} & A_1(\beta) & = & a_1+\beta\sum_{i=1}^{n_1^*}t_{i:n} + \sum_{i=n_1^*+1}^{n^*}(t_{i:n}-\tau_1+\tau_1\beta),\nonumber \\
{} & A_2(\beta,\,\theta_2) & = & a_0-\sum_{i=1}^{n_1^*}log(1-e^{-\beta\theta_2
t_{i:n}})-\sum_{i=n_1^*+1}^{n^*}log(1-e^{-\theta_2(t_{i:n}-\tau_1+\tau_1\beta)}), \nonumber \\
{} & A_3(\beta,\,\theta_2,\,\alpha) & = & 1-\{1-e^{-\theta_2(\tau^*-\tau_1+\tau_1\beta)}\}^{\alpha}.
\nonumber
\end{array}
\end{align}
Here $\tau^*=\tau_2$ and in case (a) $n_1^*=0,$ $n_2^*=n_2,$ in case (b) $n_1^*=n_1,$ $n_2^*=n_2,$ and in case (c) $n_1^*=n,$ $n_2^*=0.$

The Bayes estimate and the associated HPD credible interval of any parametric function of $(\beta,\theta_2,\alpha)$ can be obtain 
using the same algorithm as discussed in case of complete data.

\subsection{\sc Type-II Censoring}
In this censoring scheme the life testing experiment is terminated when the $rth$ (prefixed number) failure occurs, i.e, the total number of failure is fixed but the
termination time of the experiment is random. Available data under this censoring scheme is one of the forms. \newline
$(a)$ \quad $\{\tau_1<t_{1:n}<...<t_{r:n}\},$ \newline 
$(b)$ \quad $\{t_{1:n}<t_{2:n}<...<t_{n_1:n}<\tau_1<t_{n_1+1:n}<...<t_{r:n} \}, n_1<r,$ \newline 
$(c)$ \quad $\{t_{1:n}<t_{2:n}<...<t_{r:n}<\tau_1<\tau_2$\}. \newline 
Based on Type-II censored data, the posterior analysis is same as that of Type-I censoring scheme with $\tau^* = t_{r:n},$ $n^*=r$ and in case (a) $n_1^*=0,$
$n_2^*=r$; in case (b) $n_1^*=n_1,$ $n_2^*=r-n_1;$ in case (c) $n_1^*=r,$ $n_2^*=0.$ All other expressions and the following analysis are same as the Type-I 
censoring scheme.

\subsection{\sc Type-I Hybrid Censoring}
The termination time in Type-I HCS is $\tau^* = min\{t_{r:n},\tau_2\}.$ Let $n_1$ and $n_2$ be the number of failures under stress level
$S_1$ and $S_2$, respectively. Available data under this censoring scheme is one of the forms \newline
$(a)$ \quad $\{\tau_1<t_{1:n}<...<t_{r:n}\}$ if $t_{r:n}\leq\tau_2,$ \newline
$(b)$ \quad $\{t_{1:n}<t_{2:n}<...<t_{n_1:n}<\tau_1<t_{n_1+1:n}<...<t_{r:n} \}$  if $t_{r:n}<\tau_2, n_1<r, $\newline 
$(c)$ \quad $\{t_{1:n}<t_{2:n}<...<t_{r:n}<\tau_1<\tau_2$\} if $t_{r:n}<\tau_1,$\newline 
$(d)$ \quad $\{\tau_1<t_{1:n}<...<t_{n_2:n}<\tau_2\}$ if $t_{r:n}>\tau_2,$ \newline
$(e)$ \quad $\{t_{1:n}<...<t_{n_1:n}<\tau_1<t_{n_1+1:n}<...<t_{n_1+n_2:n}<\tau_2\}$ if $t_{r:n}>\tau_2, n_1<r,$ \newline
$(f)$ \quad $\{t_{1:n}<...<t_{n_1:n}<\tau_1<\tau_2\}$ if $t_{r:n}>\tau_2.$ \newline
Based on Type-I Hybrid censored data, the posterior analysis is same as that of Type-I censoring scheme with, for case (a) $n_1^*=0, n_2^* = r,$
for case (b) $n_1^* = n_1, n_2^* = r-n_1,$ for case (c) $n_1^* = r, n_2^* = 0,$ for case (d) $n_1^* = 0, n_2^* = n_2,$ for case (e)
$n_1^* = n_1, n_2^* = n_2,$ and for case (f) $n_1^* = n_1, n_2^* = 0.$ All other expressions and the following analysis are same 
as the Type-I 
censoring scheme.

\subsection{\sc Type-II Hybrid Censoring}
In Type-II HCS the experiment is terminated at $\tau^* = max\{t_{r:n},\tau_2\}.$ In this case the experimental time and the number of
failures both are random but it ensures at least $r$ failures from the experiment. Let $n_1$ and $n_2$ be the number of failures under stress level
$S_1$ and $S_2$, respectively. Available data under this censoring scheme is one of the
forms \newline
$(a)$ \quad $\{\tau_1<t_{1:n}<...<t_{r:n}\}$ if $t_{r:n}\geq\tau_2,$ \newline
$(b)$ \quad $\{t_{1:n}<t_{2:n}<...<t_{n_1:n}<\tau_1<t_{n_1+1:n}<...<t_{r:n} \}$  if $t_{r:n}\geq\tau_2, n_1<r, $\newline
$(c)$ \quad $\{\tau_1<t_{1:n}<...<t_{n_2:n}<\tau_2\}$ if $t_{r:n}<\tau_2,$ \newline
$(d)$ \quad $\{t_{1:n}<...<t_{n_1:n}<\tau_1<t_{n_1+1:n}<...<t_{n_1+n_2:n}<\tau_2\}$ if $t_{r:n}<\tau_2, n_1<r,$ \newline
$(e)$ \quad $\{t_{1:n}<...<t_{n_1:n}<\tau_1<\tau_2\}$ if $t_{r:n}<\tau_2.$ \newline
Based on the Type-II Hybrid censored data, the posterior analysis is same as that of the Type-I censoring scheme with, for case (a) $n_1^*=0, n_2^* = r,$
for case (b) $n_1^* = n_1, n_2^* = r-n_1,$ for case (c) $n_1^* = 0, n_2^* = n_2,$ for case (d) $n_1^* = n_1, n_2^* = n_2,$ for case (e)
$n_1^* = n_1, n_2^* = 0.$  All other expressions and the following analysis are same as the Type-II censoring scheme.


\section{\sc Simulation and Data Analysis}

\subsection{\sc Simulation}
In this section first we perform some simulation experiments on complete data to evaluate the performances of proposed method.   
In this simulation study we consider almost non-informative priors on $\alpha,$ $\beta$ and $\theta_2,$ i.e., $a_0=0.0001,$
$b_0=0.0001,$ $a_1=0.0001,$ $b_1=0.0001,$ $a_2=1$ and $b_2=1$ as suggested by \citet{B:PC2006}.  Results are obtained on 
$5000$ replications with $N$ = 15000.  The  Bayes estimates and the associated mean square errors (MSEs) for different parameter 
values are obtained and they are presented in Tables  
\ref{tab:AE1}, \ref{tab:AE2} and \ref{tab:AE3}.  As expected, the MSEs of Bayes estimates decrease as $n$ increases.  
Also we provide the $95\%$ symmetric and HPD CRI of the different parameters in Tables \ref{tab:CRI1}, 
\ref{tab:CRI2} and \ref{tab:CRI3}.  It has been observed that
most of the cases average estimates (AE) are overestimated for all the parameters.   Hence, we also consider the left sided
CRIs in simulation study.

\begin{table}[!ht]\scriptsize
\caption{AEs and MSEs of $\alpha$, $\theta_1$, and $\theta_2$ based on $5000$ simulations with $\alpha=0.6$,
$\theta_1=0.1$, and $\theta_2=0.2$ for different values of $n$ and $\tau$.}
\centering
\begin{tabular}{*{8}{c}}
\toprule
\multicolumn{2}{c}{} & \multicolumn{2}{c}{$\alpha$} & \multicolumn{2}{c}{$\theta_1$} &
\multicolumn{2}{c}{$\theta_2$}\\
\cmidrule(lr){3-4}\cmidrule(lr){5-6}\cmidrule(lr){7-8}
\multicolumn{1}{c}{$n$} & \multicolumn{1}{c}{$\tau$}  & \multicolumn{1}{c}{AE} & \multicolumn{1}{c}{MSE} &
\multicolumn{1}{c}{AE} & \multicolumn{1}{c}{MSE} & \multicolumn{1}{c}{AE} & \multicolumn{1}{c}{MSE} \\
\midrule
10 &  5 & 0.7598 & 0.2177 & 0.1285 & 0.0061 & 0.2811 & 0.0758\\
{} &  7 & 0.7669 & 0.2552 & 0.1246 & 0.0051 & 0.3040 & 0.1416\\ 
{} &  9 & 0.7639 & 0.2163 & 0.1201 & 0.0041 & 0.3437 & 0.5255\\[3mm]
20 &  5 & 0.6772 & 0.0633 & 0.1180 & 0.0027 & 0.2315 & 0.0102\\
{} &  7 & 0.6745 & 0.0554 & 0.1157 & 0.0024 & 0.2394 & 0.0155\\
{} &  9 & 0.6711 & 0.0545 & 0.1144 & 0.0021 & 0.2531 & 0.0272\\[3mm]
30 &  5 & 0.6544 & 0.0331 & 0.1155 & 0.0018 & 0.2218 & 0.0056\\
{} &  7 & 0.6483 & 0.0316 & 0.1125 & 0.0014 & 0.2207 & 0.0059\\
{} &  9 & 0.6522 & 0.0294 & 0.1115 & 0.0013 & 0.2306 & 0.0229\\[3mm]
40 &  5 & 0.6491 & 0.0235 & 0.1151 & 0.0015 & 0.2172 & 0.0038\\
{} &  7 & 0.6427 & 0.0201 & 0.1119 & 0.0012 & 0.2161 & 0.0045\\
{} &  9 & 0.6421 & 0.0201 & 0.1113 & 0.0010 & 0.2217 & 0.0062\\[3mm]
50 &  5 & 0.6424 & 0.0173 & 0.1137 & 0.0012 & 0.2123 & 0.0026\\
{} &  7 & 0.6406 & 0.0162 & 0.1127 & 0.0010 & 0.2160 & 0.0034\\
{} &  9 & 0.6380 & 0.0152 & 0.1114 & 0.0009 & 0.2184 & 0.0045\\
\bottomrule
\end{tabular}
\label{tab:AE1}
\end{table}

\begin{table}[!ht]\scriptsize
\caption{AEs and MSEs of $\alpha$, $\theta_1$, and $\theta_2$ based on $5000$ simulations with $\alpha=1.0$,
$\theta_1=0.1$, and $\theta_2=0.2$ for different values of $n$ and $\tau$.}
\centering
\begin{tabular}{*{8}{c}}
\toprule
\multicolumn{2}{c}{} & \multicolumn{2}{c}{$\alpha$} & \multicolumn{2}{c}{$\theta_1$} &
\multicolumn{2}{c}{$\theta_2$}\\
\cmidrule(lr){3-4}\cmidrule(lr){5-6}\cmidrule(lr){7-8}
\multicolumn{1}{c}{$n$} & \multicolumn{1}{c}{$\tau$}  & \multicolumn{1}{c}{AE} & \multicolumn{1}{c}{MSE} &
\multicolumn{1}{c}{AE} & \multicolumn{1}{c}{MSE} & \multicolumn{1}{c}{AE} & \multicolumn{1}{c}{MSE} \\
\midrule
10 &  5 & 1.3876 & 0.9952 & 0.1245 & 0.0048 & 0.2438 & 0.0214\\
{} &  7 & 1.3850 & 1.0298 & 0.1222 & 0.0042 & 0.2574 & 0.0417\\ 
{} &  9 & 1.3498 & 0.8898 & 0.1183 & 0.0036 & 0.2710 & 0.1230\\[3mm]
20 &  5 & 1.1687 & 0.2340 & 0.1148 & 0.0022 & 0.2152 & 0.0049\\
{} &  7 & 1.1596 & 0.2240 & 0.1130 & 0.0019 & 0.2204 & 0.0065\\
{} &  9 & 1.1377 & 0.2006 & 0.1098 & 0.0016 & 0.2250 & 0.0105\\[3mm]
30 &  5 & 1.1179 & 0.1374 & 0.1125 & 0.0016 & 0.2084 & 0.0029\\
{} &  7 & 1.1159 & 0.1330 & 0.1099 & 0.0013 & 0.2093 & 0.0031\\
{} &  9 & 1.1149 & 0.1277 & 0.1090 & 0.0012 & 0.2126 & 0.0042\\[3mm]
40 &  5 & 1.1024 & 0.0981 & 0.1117 & 0.0014 & 0.2059 & 0.0021\\
{} &  7 & 1.0934 & 0.0890 & 0.1091 & 0.0010 & 0.2060 & 0.0022\\
{} &  9 & 1.0778 & 0.0781 & 0.1068 & 0.0008 & 0.2070 & 0.0027\\[3mm]
50 &  5 & 1.0864 & 0.0746 & 0.1108 & 0.0012 & 0.2043 & 0.0016\\
{} &  7 & 1.0739 & 0.0653 & 0.1080 & 0.0009 & 0.2050 & 0.0018\\
{} &  9 & 1.0676 & 0.0633 & 0.1067 & 0.0007 & 0.2052 & 0.0022\\
\bottomrule
\end{tabular}
\label{tab:AE2}
\end{table}

\begin{table}[!ht]\scriptsize
\caption{AEs and MSEs of $\alpha$, $\theta_1$, and $\theta_2$ based on $5000$ simulations with $\alpha=1.5$,
$\theta_1=0.1$, and $\theta_2=0.2$ for different values of $n$ and $\tau$.}
\centering
\begin{tabular}{*{8}{c}}
\toprule
\multicolumn{2}{c}{} & \multicolumn{2}{c}{$\alpha$} & \multicolumn{2}{c}{$\theta_1$} &
\multicolumn{2}{c}{$\theta_2$}\\
\cmidrule(lr){3-4}\cmidrule(lr){5-6}\cmidrule(lr){7-8}
\multicolumn{1}{c}{$n$} & \multicolumn{1}{c}{$\tau$}  & \multicolumn{1}{c}{AE} & \multicolumn{1}{c}{MSE} &
\multicolumn{1}{c}{AE} & \multicolumn{1}{c}{MSE} & \multicolumn{1}{c}{AE} & \multicolumn{1}{c}{MSE} \\
\midrule
10 &  5 & 2.0073 & 1.6746 & 0.1180 & 0.0030 & 0.2228 & 0.0089\\
{} &  7 & 2.0745 & 1.9250 & 0.1167 & 0.0030 & 0.2309 & 0.0113\\ 
{} &  9 & 2.0925 & 2.1423 & 0.1142 & 0.0026 & 0.2395 & 0.0380\\[3mm]
20 &  5 & 1.7279 & 0.4353 & 0.1100 & 0.0016 & 0.2081 & 0.0032\\
{} &  7 & 1.7431 & 0.4939 & 0.1077 & 0.0014 & 0.2090 & 0.0036\\
{} &  9 & 1.7316 & 0.4566 & 0.1080 & 0.0013 & 0.2168 & 0.0047\\[3mm]
30 &  5 & 1.6468 & 0.2595 & 0.1052 & 0.0011 & 0.2020 & 0.0018\\
{} &  7 & 1.6424 & 0.2727 & 0.1050 & 0.0010 & 0.2023 & 0.0019\\
{} &  9 & 1.6461 & 0.2619 & 0.1057 & 0.0009 & 0.2065 & 0.0026\\[3mm]
40 &  5 & 1.6035 & 0.1714 & 0.1048 & 0.0010 & 0.2003 & 0.0014\\
{} &  7 & 1.5871 & 0.1629 & 0.1033 & 0.0008 & 0.1986 & 0.0014\\
{} &  9 & 1.5937 & 0.1617 & 0.1029 & 0.0007 & 0.2014 & 0.0018\\[3mm]
50 &  5 & 1.5662 & 0.1269 & 0.1027 & 0.0009 & 0.1980 & 0.0011\\
{} &  7 & 1.5718 & 0.1320 & 0.1023 & 0.0007 & 0.1982 & 0.0011\\
{} &  9 & 1.5667 & 0.1205 & 0.1022 & 0.0006 & 0.2000 & 0.0014\\
\bottomrule
\end{tabular}
\label{tab:AE3}
\end{table}


\begin{landscape}
\begin{table}[!ht]\scriptsize
\caption{CPs and ALs of 95\% CRI for $\alpha,$ $\theta_1$ and $\theta_2$ based on $5000$ simulations with $\alpha=0.6$, $\theta_1=0.1$,
and $\theta_2=0.2$ for different values of $n$ and $\tau$.}
\centering
\begin{tabular}{*{20}{c}}
\toprule
\multicolumn{2}{c}{} & \multicolumn{6}{c}{$\alpha$} & \multicolumn{6}{c}{$\theta_1$} & \multicolumn{6}{c}{$\theta_2$} \\
\cmidrule(lr){3-8}\cmidrule(lr){9-14}\cmidrule(lr){15-20}
\multicolumn{2}{c}{} & \multicolumn{2}{c}{Left CRI} & \multicolumn{2}{c}{Symmetric CRI} &  \multicolumn{2}{c}{HPD CRI}
& \multicolumn{2}{c}{Left CRI} & \multicolumn{2}{c}{Symmetric CRI} &  \multicolumn{2}{c}{HPD CRI}
& \multicolumn{2}{c}{Left CRI} & \multicolumn{2}{c}{Symmetric CRI} &  \multicolumn{2}{c}{HPD CRI}\\
\cmidrule(lr){3-4}\cmidrule(lr){5-6}\cmidrule(lr){7-8}\cmidrule(lr){9-10}\cmidrule(lr){11-12}\cmidrule(lr){13-14}\cmidrule(lr){15-16}\cmidrule(lr){17-18}\cmidrule(lr){19-20}  
\multicolumn{1}{c}{$n$} & \multicolumn{1}{c}{$\tau$}  & \multicolumn{1}{c}{CP} & \multicolumn{1}{c}{AL} &
\multicolumn{1}{c}{CP} & \multicolumn{1}{c}{AL} & \multicolumn{1}{c}{CP} & \multicolumn{1}{c}{AL} &
\multicolumn{1}{c}{CP} & \multicolumn{1}{c}{AL} & \multicolumn{1}{c}{CP} & \multicolumn{1}{c}{AL} &
\multicolumn{1}{c}{CP} & \multicolumn{1}{c}{AL} & \multicolumn{1}{c}{CP} & \multicolumn{1}{c}{AL} &
\multicolumn{1}{c}{CP} & \multicolumn{1}{c}{AL} & \multicolumn{1}{c}{CP} & \multicolumn{1}{c}{AL}\\
\midrule
10 &  5 & 95.36 & 1.3426 & 96.04 & 1.3458 & 95.06 & 1.2343 & 97.66 & 0.2657 & 97.26 & 0.2728 & 96.40 & 0.2504 & 95.70 & 0.5171 & 94.94 & 0.5722 & 95.18 & 0.5090\\
{} &  7 & 94.88 & 1.3446 & 95.54 & 1.3312 & 94.52 & 1.2265 & 97.48 & 0.2489 & 97.12 & 0.2497 & 95.86 & 0.2298 & 95.58 & 0.6246 & 95.36 & 0.7248 & 95.18 & 0.6178\\  
{} &  9 & 95.14 & 1.3254 & 95.42 & 1.3088 & 94.82 & 1.2037 & 96.96 & 0.2354 & 97.54 & 0.2331 & 95.82 & 0.2144 & 96.10 & 0.8630 & 96.20 & 1.0492 & 95.90 & 0.8578\\[3mm]
20 &  5 & 95.82 & 0.9474 & 96.42 & 0.8134 & 95.84 & 0.7687 & 97.60 & 0.2114 & 96.66 & 0.1927 & 96.06 & 0.1813 & 95.48 & 0.2932 & 94.58 & 0.3044 & 94.10 & 0.2824\\
{} &  7 & 96.18 & 0.9395 & 95.94 & 0.7905 & 96.16 & 0.7485 & 97.40 & 0.2003 & 95.58 & 0.1749 & 95.22 & 0.1646 & 95.30 & 0.3314 & 94.78 & 0.3510 & 94.84 & 0.3220\\
{} &  9 & 96.00 & 0.9388 & 95.50 & 0.7772 & 95.78 & 0.7377 & 97.54 & 0.1954 & 95.36 & 0.1650 & 95.28 & 0.1550 & 95.86 & 0.4657 & 95.68 & 0.5419 & 95.56 & 0.4578\\[3mm]
30 &  5 & 95.72 & 0.8218 & 95.64 & 0.6303 & 95.44 & 0.5992 & 97.68 & 0.1918 & 95.34 & 0.1590 & 95.00 & 0.1506 & 94.86 & 0.2311 & 93.64 & 0.2345 & 92.96 & 0.2192\\
{} &  7 & 95.80 & 0.8124 & 95.28 & 0.6040 & 95.24 & 0.5758 & 97.22 & 0.1799 & 95.10 & 0.1406 & 94.36 & 0.1330 & 95.36 & 0.2476 & 95.14 & 0.2550 & 94.42 & 0.2371\\
{} &  9 & 95.86 & 0.8162 & 95.34 & 0.5979 & 95.28 & 0.5701 & 97.08 & 0.1743 & 94.10 & 0.1310 & 93.80 & 0.1234 & 95.40 & 0.2860 & 95.80 & 0.3010 & 94.82 & 0.2769\\[3mm]
40 &  5 & 96.26 & 0.7632 & 95.62 & 0.5341 & 95.68 & 0.5085 & 97.26 & 0.1809 & 94.20 & 0.1381 & 93.50 & 0.1310 & 94.88 & 0.1961 & 92.98 & 0.1952 & 92.10 & 0.1833\\
{} &  7 & 96.22 & 0.7558 & 95.74 & 0.5111 & 95.78 & 0.4877 & 97.42 & 0.1699 & 93.26 & 0.1211 & 93.16 & 0.1147 & 95.06 & 0.2123 & 94.50 & 0.2145 & 93.66 & 0.2010\\
{} &  9 & 96.16 & 0.7543 & 95.30 & 0.4993 & 95.08 & 0.4765 & 97.10 & 0.1647 & 92.52 & 0.1115 & 92.32 & 0.1050 & 94.78 & 0.2383 & 95.20 & 0.2454 & 93.66 & 0.2284\\[3mm]
50 &  5 & 96.56 & 0.7210 & 95.56 & 0.4658 & 95.48 & 0.4432 & 97.62 & 0.1721 & 93.14 & 0.1225 & 92.70 & 0.1162 & 94.98 & 0.1707 & 92.82 & 0.1677 & 91.72 & 0.1578\\
{} &  7 & 96.34 & 0.7211 & 95.10 & 0.4487 & 95.16 & 0.4276 & 97.88 & 0.1646 & 91.18 & 0.1084 & 91.16 & 0.1023 & 95.38 & 0.1922 & 94.26 & 0.1909 & 93.32 & 0.1793\\
{} &  9 & 96.26 & 0.7207 & 95.28 & 0.4392 & 95.14 & 0.4184 & 97.36 & 0.1583 & 90.82 & 0.0983 & 90.62 & 0.0924 & 95.00 & 0.2098 & 94.90 & 0.2123 & 93.64 & 0.1981\\
\bottomrule
\end{tabular}
\label{tab:CRI1}
\end{table}
\end{landscape}


\begin{landscape}
\begin{table}[!ht]\scriptsize
\caption{CPs and ALs of 95\% CRI for $\alpha,$ $\theta_1$ and $\theta_2$ based on $5000$ simulations with $\alpha=1$, $\theta_1=0.1$,
and $\theta_2=0.2$ for different values of $n$ and $\tau$.}
\centering
\begin{tabular}{*{20}{c}}
\toprule
\multicolumn{2}{c}{} & \multicolumn{6}{c}{$\alpha$} & \multicolumn{6}{c}{$\theta_1$} & \multicolumn{6}{c}{$\theta_2$} \\
\cmidrule(lr){3-8}\cmidrule(lr){9-14}\cmidrule(lr){15-20}
\multicolumn{2}{c}{} & \multicolumn{2}{c}{Left CRI} & \multicolumn{2}{c}{Symmetric CRI} &  \multicolumn{2}{c}{HPD CRI}
& \multicolumn{2}{c}{Left CRI} & \multicolumn{2}{c}{Symmetric CRI} &  \multicolumn{2}{c}{HPD CRI}
& \multicolumn{2}{c}{Left CRI} & \multicolumn{2}{c}{Symmetric CRI} &  \multicolumn{2}{c}{HPD CRI}\\
\cmidrule(lr){3-4}\cmidrule(lr){5-6}\cmidrule(lr){7-8}\cmidrule(lr){9-10}\cmidrule(lr){11-12}\cmidrule(lr){13-14}\cmidrule(lr){15-16}\cmidrule(lr){17-18}\cmidrule(lr){19-20}  
\multicolumn{1}{c}{$n$} & \multicolumn{1}{c}{$\tau$}  & \multicolumn{1}{c}{CP} & \multicolumn{1}{c}{AL} &
\multicolumn{1}{c}{CP} & \multicolumn{1}{c}{AL} & \multicolumn{1}{c}{CP} & \multicolumn{1}{c}{AL} &
\multicolumn{1}{c}{CP} & \multicolumn{1}{c}{AL} & \multicolumn{1}{c}{CP} & \multicolumn{1}{c}{AL} &
\multicolumn{1}{c}{CP} & \multicolumn{1}{c}{AL} & \multicolumn{1}{c}{CP} & \multicolumn{1}{c}{AL} &
\multicolumn{1}{c}{CP} & \multicolumn{1}{c}{AL} & \multicolumn{1}{c}{CP} & \multicolumn{1}{c}{AL}\\
\midrule
10 &  5 & 96.08 & 2.7174 & 97.08 & 2.8248 & 95.30 & 2.5520  & 98.76 & 0.2508 & 98.54 & 0.2558 & 96.82 & 0.2368  & 95.68 & 0.3918 & 95.14 & 0.3921 & 93.82 & 0.3673\\
{} &  7 & 95.70 & 2.6620 & 96.80 & 2.7352 & 94.58 & 2.4778  & 97.96 & 0.2359 & 97.66 & 0.2337 & 95.24 & 0.2178  & 96.38 & 0.4433 & 95.34 & 0.4586 & 95.20 & 0.4208\\ 
{} &  9 & 95.06 & 2.5851 & 95.72 & 2.6204 & 93.48 & 2.3878  & 96.84 & 0.2218 & 96.94 & 0.2153 & 94.08 & 0.2010  & 96.12 & 0.6773 & 95.70 & 0.8150 & 95.44 & 0.6574\\[3mm]
20 &  5 & 96.38 & 1.8108 & 97.34 & 1.6592 & 95.50 & 1.5566  & 98.40 & 0.2051 & 97.92 & 0.1907 & 96.08 & 0.1807  & 94.96 & 0.2664 & 95.18 & 0.2441 & 93.78 & 0.2339\\
{} &  7 & 95.32 & 1.7860 & 96.12 & 1.5936 & 93.84 & 1.4995  & 97.40 & 0.1927 & 97.40 & 0.1708 & 94.68 & 0.1626  & 94.66 & 0.2880 & 95.02 & 0.2675 & 94.04 & 0.2558\\
{} &  9 & 94.80 & 1.7307 & 95.80 & 1.5116 & 93.70 & 1.4269  & 97.28 & 0.1819 & 96.74 & 0.1558 & 95.02 & 0.1486  & 94.98 & 0.3135 & 95.46 & 0.2999 & 94.54 & 0.2844\\[3mm]
30 &  5 & 95.96 & 1.5564 & 96.84 & 1.3035 & 94.76 & 1.2371  & 97.86 & 0.1883 & 97.68 & 0.1632 & 95.12 & 0.1557  & 94.18 & 0.2225 & 95.10 & 0.1930 & 93.32 & 0.1862\\
{} &  7 & 95.84 & 1.5466 & 95.64 & 1.2462 & 93.74 & 1.1889  & 97.34 & 0.1756 & 97.28 & 0.1434 & 94.54 & 0.1373  & 94.86 & 0.2358 & 95.56 & 0.2070 & 94.80 & 0.1998\\
{} &  9 & 95.90 & 1.5414 & 94.96 & 1.2094 & 93.16 & 1.1568  & 96.90 & 0.1684 & 96.10 & 0.1308 & 93.76 & 0.1255  & 94.54 & 0.2540 & 95.86 & 0.2285 & 94.62 & 0.2199\\[3mm]
40 &  5 & 95.82 & 1.4405 & 96.26 & 1.1248 & 93.98 & 1.0746  & 97.56 & 0.1792 & 97.08 & 0.1467 & 93.98 & 0.1403  & 94.32 & 0.1985 & 94.66 & 0.1651 & 93.64 & 0.1597\\
{} &  7 & 96.26 & 1.4271 & 96.06 & 1.0664 & 94.02 & 1.0233  & 97.16 & 0.1671 & 96.58 & 0.1280 & 93.70 & 0.1228  & 94.36 & 0.2107 & 95.52 & 0.1774 & 94.26 & 0.1721\\
{} &  9 & 96.36 & 1.3962 & 96.14 & 1.0107 & 94.62 & 0.9721  & 97.30 & 0.1586 & 96.84 & 0.1149 & 94.56 & 0.1106  & 94.10 & 0.2242 & 95.62 & 0.1942 & 94.64 & 0.1883\\[3mm]
50 &  5 & 96.10 & 1.3578 & 96.04 & 1.0029 & 94.30 & 0.9606  & 97.32 & 0.1724 & 96.72 & 0.1351 & 93.66 & 0.1293  & 94.12 & 0.1823 & 94.86 & 0.1467 & 93.96 & 0.1423\\
{} &  7 & 96.30 & 1.3421 & 96.26 & 0.9439 & 94.54 & 0.9073  & 97.18 & 0.1607 & 96.50 & 0.1166 & 93.48 & 0.1121  & 94.42 & 0.1950 & 95.12 & 0.1588 & 94.14 & 0.1546\\
{} &  9 & 95.88 & 1.3281 & 95.44 & 0.8982 & 93.56 & 0.8659  & 96.74 & 0.1534 & 95.90 & 0.1042 & 93.30 & 0.1004  & 93.96 & 0.2071 & 95.64 & 0.1738 & 94.22 & 0.1690\\
\bottomrule
\end{tabular}
\label{tab:CRI2}
\end{table}
\end{landscape}


\begin{landscape}
\begin{table}[!ht]\scriptsize
\caption{CPs and ALs of 95\% CRI for $\alpha,$ $\theta_1$ and $\theta_2$ based on $5000$ simulations with $\alpha=1.5$, $\theta_1=0.1$,
and $\theta_2=0.2$ for different values of $n$ and $\tau$.}
\centering
\begin{tabular}{*{20}{c}}
\toprule
\multicolumn{2}{c}{} & \multicolumn{6}{c}{$\alpha$} & \multicolumn{6}{c}{$\theta_1$} & \multicolumn{6}{c}{$\theta_2$} \\
\cmidrule(lr){3-8}\cmidrule(lr){9-14}\cmidrule(lr){15-20}
\multicolumn{2}{c}{} & \multicolumn{2}{c}{Left CRI} & \multicolumn{2}{c}{Symmetric CRI} &  \multicolumn{2}{c}{HPD CRI}
& \multicolumn{2}{c}{Left CRI} & \multicolumn{2}{c}{Symmetric CRI} &  \multicolumn{2}{c}{HPD CRI}
& \multicolumn{2}{c}{Left CRI} & \multicolumn{2}{c}{Symmetric CRI} &  \multicolumn{2}{c}{HPD CRI}\\
\cmidrule(lr){3-4}\cmidrule(lr){5-6}\cmidrule(lr){7-8}\cmidrule(lr){9-10}\cmidrule(lr){11-12}\cmidrule(lr){13-14}\cmidrule(lr){15-16}\cmidrule(lr){17-18}\cmidrule(lr){19-20}  
\multicolumn{1}{c}{$n$} & \multicolumn{1}{c}{$\tau$}  & \multicolumn{1}{c}{CP} & \multicolumn{1}{c}{AL} &
\multicolumn{1}{c}{CP} & \multicolumn{1}{c}{AL} & \multicolumn{1}{c}{CP} & \multicolumn{1}{c}{AL} &
\multicolumn{1}{c}{CP} & \multicolumn{1}{c}{AL} & \multicolumn{1}{c}{CP} & \multicolumn{1}{c}{AL} &
\multicolumn{1}{c}{CP} & \multicolumn{1}{c}{AL} & \multicolumn{1}{c}{CP} & \multicolumn{1}{c}{AL} &
\multicolumn{1}{c}{CP} & \multicolumn{1}{c}{AL} & \multicolumn{1}{c}{CP} & \multicolumn{1}{c}{AL}\\
\midrule
10 &  5 & 96.14 & 4.0881 & 97.82 & 4.2513 & 96.02 & 3.8536  & 99.10 & 0.2352 & 99.28 & 0.2377 & 98.14 & 0.2217  & 94.80 & 0.3361 & 95.02 & 0.3148 & 93.36 & 0.2992\\
{} &  7 & 96.36 & 4.1680 & 97.86 & 4.3001 & 95.54 & 3.8968  & 98.62 & 0.2207 & 98.78 & 0.2161 & 96.52 & 0.2019  & 95.60 & 0.3597 & 95.72 & 0.3398 & 94.28 & 0.3215\\ 
{} &  9 & 96.10 & 4.1478 & 97.00 & 4.2346 & 95.00 & 3.8512  & 98.08 & 0.2083 & 97.96 & 0.1984 & 95.40 & 0.1860  & 95.38 & 0.3998 & 95.62 & 0.3933 & 94.22 & 0.3629\\[3mm]
20 &  5 & 96.04 & 2.7793 & 97.86 & 2.5945 & 95.84 & 2.4207  & 98.24 & 0.1944 & 98.90 & 0.1802 & 96.82 & 0.1701  & 94.04 & 0.2493 & 95.42 & 0.2049 & 93.90 & 0.1969\\
{} &  7 & 96.62 & 2.7795 & 97.74 & 2.5297 & 95.82 & 2.3624  & 97.90 & 0.1799 & 98.30 & 0.1579 & 95.64 & 0.1491  & 94.18 & 0.2603 & 94.76 & 0.2145 & 93.40 & 0.2063\\
{} &  9 & 96.10 & 2.7503 & 97.20 & 2.4378 & 94.50 & 2.2821  & 97.20 & 0.1730 & 97.70 & 0.1442 & 94.70 & 0.1363  & 94.40 & 0.2834 & 95.80 & 0.2392 & 94.70 & 0.2294\\[3mm]
30 &  5 & 97.20 & 2.3413 & 98.10 & 2.0186 & 96.40 & 1.8986  & 98.60 & 0.1744 & 99.10 & 0.1528 & 97.60 & 0.1441  & 93.60 & 0.2133 & 94.40 & 0.1608 & 92.50 & 0.1545 \\
{} &  7 & 95.60 & 2.3496 & 97.20 & 1.9468 & 94.70 & 1.8403  & 97.60 & 0.1634 & 98.10 & 0.1325 & 95.10 & 0.1253  & 92.20 & 0.2220 & 94.50 & 0.1691 & 93.30 & 0.1628 \\
{} &  9 & 95.50 & 2.3318 & 96.50 & 1.8576 & 93.30 & 1.7544  & 98.00 & 0.1571 & 95.80 & 0.1184 & 93.50 & 0.1119  & 93.70 & 0.2369 & 94.50 & 0.1842 & 93.30 & 0.1774 \\[3mm]
40 &  5 & 96.80 & 2.1207 & 97.80 & 1.6984 & 95.90 & 1.6066  & 97.60 & 0.1647 & 98.10 & 0.1356 & 94.90 & 0.1280  & 90.50 & 0.1923 & 92.20 & 0.1355 & 90.50 & 0.1301 \\
{} &  7 & 96.50 & 2.1081 & 97.40 & 1.6193 & 94.60 & 1.5290  & 97.00 & 0.1536 & 97.40 & 0.1162 & 94.40 & 0.1098  & 92.40 & 0.2004 & 94.10 & 0.1427 & 93.30 & 0.1377 \\
{} &  9 & 95.46 & 2.1156 & 96.72 & 1.5526 & 94.06 & 1.4738  & 96.56 & 0.1467 & 96.80 & 0.1025 & 93.58 & 0.0968  & 92.40 & 0.2136 & 94.54 & 0.1563 & 92.90 & 0.1508\\[3mm]
50 &  5 & 96.84 & 1.9516 & 97.96 & 1.4749 & 96.18 & 1.3982  & 97.14 & 0.1560 & 97.90 & 0.1228 & 95.28 & 0.1159  & 91.32 & 0.1780 & 92.76 & 0.1186 & 90.92 & 0.1138\\
{} &  7 & 96.52 & 1.9784 & 97.20 & 1.4123 & 94.86 & 1.3377  & 97.06 & 0.1462 & 96.90 & 0.1039 & 93.78 & 0.0980  & 90.96 & 0.1873 & 93.28 & 0.1259 & 91.46 & 0.1212\\
{} &  9 & 95.94 & 1.9781 & 96.38 & 1.3479 & 94.36 & 1.2799  & 96.46 & 0.1404 & 96.60 & 0.0913 & 93.24 & 0.0862  & 91.66 & 0.1991 & 93.78 & 0.1384 & 92.48 & 0.1335\\
\bottomrule
\end{tabular}
\label{tab:CRI3}
\end{table}
\end{landscape}

We have further performed some simulation experiments based on Type-I and Type-II censored data.  We have taken the same parameter
values and the priors.  The order restricted Bayes estimates and the associated MSEs 
of Type-I and Type-II censored data are presented in Tables \ref{tab:AE4} and \ref{tab:AE5}, respectively.  $95\%$ CRIs of censored 
data are provided in Tables \ref{tab:CRI4} and \ref{tab:CRI5}.  Tables \ref{tab:AE4} to \ref{tab:CRI5} are provided in the 
Appendix \ref{th:simulation}.  Censored data simulation results are very similar to that of complete data.  
In all the cases the parameter estimates are very consistent and the coverage percentages (CP) are very close to the nominal values.  Also 
average lengths (AL) of CRIs are gradually decreases as sample size increases.

\subsection{\sc Data Analysis}
\subsubsection{\sc Simulated Data Analysis}

Here we consider the analysis of two simulated data sets; one the shape parameter is less than one and other it is greater than one.
Data presented in Table \ref{tab:data1} is generated from (\ref{eq:CDF}) with $\alpha=0.6,$ $\theta_1=0.1$  $\theta_2 = 0.2,$ $n=20$ 
and $\tau_1=5.$  Artificially we have created Type-I and Type-II censored data by taking $\tau_2=8$ and $r=16$, respectively.
Prior assumptions are same as considered in simulation study.  For Type-I censored data the Bayes estimates
of $\alpha,$ $\theta_1,$ and $\theta_2$ under the squared error loss function are $0.6995, 0.1032,$ and $0.2747$,
respectively. In case of Type-II censored data Bayes estimates of $\alpha,$ $\theta_1,$ and $\theta_2$ are  $0.6244, 0.0840$
and $0.2659$ respectively. Different CRIs for both Type-I and Type-II of censoring schemes are given in Table \ref{tab:data1CRI}.

We analyze another data presented in Table \ref{tab:data2} which is generated from the (\ref{eq:CDF}) with $\alpha=1.5.$ All other
parameter values are same as the first data set. Here also we have considered Type-I and Type-II censored data.  The Bayes estimates 
of $\alpha,$ $\theta_1,$ and $\theta_2$ in Type-I censoring are $1.2787,  0.1109,$ and  $0.2269$, respectively. In Type-II
censored data Bayes estimates of $\alpha,$ $\theta_1,$ and $\theta_2$ are $1.2147,    0.1041,$ and  $0.2220$, respectively.
$90\%,$ $95\%$ and $99\%$ CRIs for both Type-I and Type-II censoring schemes are reported in Table \ref{tab:data2CRI}.

\begin{table}[!ht]\scriptsize
\caption{Type-I and Type-II censored data for analysis with $\alpha=0.6$}
\centering
\begin{tabular}{lc*{8}{c}}
\toprule
Censoring Scheme & Stress Level & \multicolumn{8}{c}{Data} \\
\midrule
Type-I and Type-II & $S_1$ & 0.0185  &  0.0763  &  1.0137  &  1.2043  &  1.3411 &   1.3968 & 2.6797  &  3.4931 \\
Type-I  &  $S_2$  &  5.1680 & 5.2476 &   5.4308 &   5.9575 &   7.2580 & 7.5416  &  7.7453 & {} \\
Type-II &  $S_2$  & 5.1680  & 5.2476 &   5.4308 &   5.9575 &   7.2580 & 7.5416  &  7.7453 &   8.0116 \\

\bottomrule
\end{tabular}
\label{tab:data1}
\end{table}

\begin{table}[!ht]\scriptsize
\caption{CRIs for the unknown parameters for data in Table \ref{tab:data1} }
\centering
\begin{tabular}{cl*{12}{c}}
\toprule
\multicolumn{2}{c}{} & \multicolumn{6}{c}{Type-I Censored data} & \multicolumn{6}{c}{Type-II Censored data} \\
\cmidrule(lr){3-8}\cmidrule(lr){9-14}
{} & {} & \multicolumn{2}{c}{Left} & \multicolumn{2}{c}{Symmetric} & \multicolumn{2}{c}{HPD} & \multicolumn{2}{c}{Left} & \multicolumn{2}{c}{Symmetric} & \multicolumn{2}{c}{HPD}  \\  
\cmidrule(lr){1-2}\cmidrule(lr){3-4}\cmidrule(lr){5-6}\cmidrule(lr){7-8}\cmidrule(lr){9-10}\cmidrule(lr){11-12}\cmidrule(lr){13-14}
Parameters & Level & LL & UL & LL & UL & LL & UL & LL & UL & LL & UL & LL & UL \\
\midrule
$\alpha$   & $90\%$ & 0.1186  &  0.9681  &  0.4207  &  1.1003  &  0.4129  &  1.0244 & 0.0852  &  0.8891  &  0.3713  &  1.0242  &  0.2727  &  0.8892   \\ 
{}         & $95\%$ & 0.1186  &  1.1003  &  0.4050  &  1.1396  &  0.4152  &  1.1396 & 0.0852  &  1.0242  &  0.2738  &  1.1271  &  0.2727  &  1.0242   \\ 
{}         & $99\%$ & 0.1186  &  1.2732  &  0.3015  &  1.3605  &  0.3015  &  1.2732 & 0.0852  &  1.2267  &  0.2738  &  1.2989  &  0.2727  &  1.2295   \\ 
\\
$\theta_1$ & $90\%$ & 0.0001  &  0.1472  &  0.0593  &  0.1676  &  0.0561  &  0.1564 & 0.0001  &  0.1323  &  0.0375  &  0.1550  &  0.0375  &  0.1362   \\ 
{}         & $95\%$ & 0.0001  &  0.1676  &  0.0561  &  0.1853  &  0.0484  &  0.1704 & 0.0001  &  0.1550  &  0.0375  &  0.1747  &  0.0375  &  0.1625   \\
{}         & $99\%$ & 0.0001  &  0.2098  &  0.0484  &  0.2255  &  0.0466  &  0.2219 & 0.0001  &  0.1983  &  0.0267  &  0.2152  &  0.0203  &  0.1996   \\
\\
$\theta_2$ & $90\%$ & 0.1041  &  0.3981  &  0.1469  &  0.4496  &  0.1234  &  0.4048 & 0.0976  &  0.3903  &  0.1397  &  0.4496  &  0.1352  &  0.4118   \\
{}         & $95\%$ & 0.1041  &  0.4496  &  0.1252  &  0.4973  &  0.1163  &  0.4535 & 0.0976  &  0.4496  &  0.1354  &  0.5042  &  0.1031  &  0.4506   \\
{}         & $99\%$ & 0.1041  &  0.5548  &  0.1113  &  0.5845  &  0.1041  &  0.5548 & 0.0976  &  0.5618  &  0.1034  &  0.5987  &  0.1031  &  0.5657   \\ 

\bottomrule
\end{tabular}
\label{tab:data1CRI}
\end{table}

\begin{table}[!ht]\scriptsize
\caption{Type-I and Type-II censored data for analysis with $\alpha=1.5$}
\centering
\begin{tabular}{lc*{7}{c}}
\toprule
Censoring Scheme & Stress Level & \multicolumn{7}{c}{Data} \\
\midrule
Type-I and Type-II & $S_1$ & 0.6277  &  0.7266  &  2.2977  &  2.8450  &  3.0599  &  3.3134 & {}  \\
Type-I  &  $S_2$ &  5.1058 &   5.4453  &  5.5445  &  6.3469  &  7.1927  &  7.2401  &  7.5872  \\
Type-II &  $S_2$  &  5.1058  &  5.4453 &   5.5445 &  6.3469  &  7.1927  &  7.2401  &  7.5872 \\ 
{} & {} & 8.0156 &   8.0383 & 10.7256  & {} & {} & {} & {}\\

\bottomrule
\end{tabular}
\label{tab:data2}
\end{table}

\begin{table}[!ht]\scriptsize
\caption{CRIs for the unknown parameters for data in Table \ref{tab:data2} }
\centering
\begin{tabular}{cl*{12}{c}}
\toprule
\multicolumn{2}{c}{} & \multicolumn{6}{c}{Type-I Censored data} & \multicolumn{6}{c}{Type-II Censored data} \\
\cmidrule(lr){3-8}\cmidrule(lr){9-14}
{} & {} & \multicolumn{2}{c}{Left} & \multicolumn{2}{c}{Symmetric} & \multicolumn{2}{c}{HPD} & \multicolumn{2}{c}{Left} & \multicolumn{2}{c}{Symmetric} & \multicolumn{2}{c}{HPD}  \\  
\cmidrule(lr){1-2}\cmidrule(lr){3-4}\cmidrule(lr){5-6}\cmidrule(lr){7-8}\cmidrule(lr){9-10}\cmidrule(lr){11-12}\cmidrule(lr){13-14}
Parameters & Level & LL & UL & LL & UL & LL & UL & LL & UL & LL & UL & LL & UL \\
\midrule
$\alpha$   & $90\%$ &   0.1246  &  1.9342  &  0.6761  &  2.1827  &  0.6668  &  2.0429 &  0.1286  &  1.8628  &  0.6028  &  2.1363  &  0.4506  &  1.8835  \\ 
{}         & $95\%$ &   0.1246  &  2.1827  &  0.6378  &  2.4342  &  0.5816  &  2.2610 &  0.1286  &  2.1363  &  0.4769  &  2.4069  &  0.4360  &  2.1514  \\ 
{}         & $99\%$ &   0.1246  &  2.7845  &  0.5143  &  3.0832  &  0.4637  &  2.8129 &  0.1286  &  2.8159  &  0.4506  &  3.1361  &  0.4248  &  2.8789  \\ 
\\
$\theta_1$ & $90\%$ &   0.0004  &  0.1805  &  0.0518  &  0.2056  &  0.0424  &  0.1830 &  0.0001  &  0.1776  &  0.0311  &  0.2037  &  0.0135  &  0.1777  \\ 
{}         & $95\%$ &   0.0004  &  0.2056  &  0.0494  &  0.2297  &  0.0424  &  0.2115 &  0.0001  &  0.2037  &  0.0210  &  0.2264  &  0.0135  &  0.2038  \\
{}         & $99\%$ &   0.0004  &  0.2572  &  0.0347  &  0.2748  &  0.0330  &  0.2579 &  0.0001  &  0.2565  &  0.0135  &  0.2789  &  0.0135  &  0.2574  \\
\\
$\theta_2$ & $90\%$ &   0.0585  &  0.3322  &  0.1061  &  0.3784  &  0.1135  &  0.3816 &  0.0690  &  0.3192  &  0.1183  &  0.3523  &  0.1013  &  0.3287  \\
{}         & $95\%$ &   0.0585  &  0.3784  &  0.0764  &  0.4248  &  0.0585  &  0.3784 &  0.0690  &  0.3523  &  0.1015  &  0.3819  &  0.1013  &  0.3731  \\
{}         & $99\%$ &   0.0585  &  0.4746  &  0.0585  &  0.5132  &  0.0585  &  0.4746 &  0.0690  &  0.4209  &  0.0774  &  0.4490  &  0.0690  &  0.4209  \\ 

\bottomrule
\end{tabular}
\label{tab:data2CRI}
\end{table}

\subsubsection{\sc Solar Lighting Device Data Set}

A simple step stress test was conducted in order to asses the reliability characteristics of a solar lighting device.  
Thirty five (35) devices are put on a life test at the normal operating temperature 293K, and then the 
stress factor temperature is changed to 353K at the time point $\tau_1$ = 5 (in hundred hours).  The experiment was terminated at the 
time point $\tau_2$ = 6 (in hundred hours).  Thirty one (31) failures occur before $\tau_2$ and among them fifteen (15) 
devices are failed at first stress and remaining sixteen (16) devices are failed at second stress level.  The data are 
presented in Table \ref{lighting}.  

\begin{table}[!ht]\scriptsize
\caption{Solar lighting device dataset    \label{lighting}}
\centering
\begin{tabular}{c*{11}{c}}
\toprule
Stress Level & \multicolumn{10}{c}{Data} \\
\midrule
$S_1$  &  0.140 & 0.783 & 1.324 & 1.582 & 1.716 & 1.794 & 1.883 & 2.293 & 2.660 & 2.674 & 2.725 \\
 {}    &  3.085 & 3.924 & 4.396 & 4.612 & 4.892  \\
$S_2$  &  5.002 & 5.022 & 5.082 & 5.112 & 5.147 & 5.238 & 5.244 & 5.247 & 5.305 & 5.337 & 5.407\\ 
{}     &  5.408 & 5.445 & 5.483 & 5.717 \\
\bottomrule
\end{tabular}
\end{table}

We analyze the solar light data set based on the assumptions that at any stress level life time of devices follow GE distribution.  We have 
obtained the order restricted Bayes estimates and different CRIs of model parameters.  The order restricted Bayes estimates of 
$\alpha$, $\theta_1$ and $\theta_2$ are respectively $1.4434$, $0.1810$ and $1.7921$.   CRIs of parameters are presented in Table \ref{tab:data3CRI}.

\begin{table}[!ht]\scriptsize
\caption{CRIs for the unknown parameters for data in Table \ref{lighting} }
\centering
\begin{tabular}{cl*{12}{c}}
\toprule
{} & {} & \multicolumn{2}{c}{Left} & \multicolumn{2}{c}{Symmetric} & \multicolumn{2}{c}{HPD}  \\  
\cmidrule(lr){1-2}\cmidrule(lr){3-4}\cmidrule(lr){5-6}\cmidrule(lr){7-8}
Parameters & Level & LL & UL & LL & UL & LL & UL  \\
\midrule
$\alpha$   & $90\%$ &   0.1948  &  2.0435  &  0.8623  &  2.2657  &  0.8249  &  2.1474  \\
{}         & $95\%$ &   0.1948  &  2.2657  &  0.8149  &  2.4153  &  0.7491  &  2.3247  \\
{}         & $99\%$ &   0.1948  &  2.6514  &  0.6694  &  2.8292  &  0.6200  &  2.6677  \\
\\
$\theta_1$ & $90\%$ &   0.0003  &  0.2609  &  0.1028  &  0.2856  &  0.0982  &  0.2654  \\
{}         & $95\%$ &   0.0003  &  0.2856  &  0.1009  &  0.3051  &  0.0989  &  0.2942  \\
{}         & $99\%$ &   0.0003  &  0.3284  &  0.0797  &  0.3480  &  0.0797  &  0.3413  \\
\\
$\theta_2$ & $90\%$ &   0.1357  &  2.4480  &  1.1284  &  2.6483  &  1.0702  &  2.5322  \\
{}         & $95\%$ &   0.1357  &  2.6483  &  1.0273  &  2.8873  &  0.9308  &  2.7295  \\
{}         & $99\%$ &   0.1357  &  3.1060  &  0.8437  &  3.2785  &  0.7681  &  3.1655  \\
\bottomrule
\end{tabular}
\label{tab:data3CRI}
\end{table}

Now one natural question whether the GE distribution fits the data set or not.  We have used the 
Kolmogorov-Smirnov (K-S) statistic, which quantifies the distance between the  
empirical distribution function of the data set and the cumulative distribution function of the fitted distribution function, 
for that purpose.  The K-S distance and associated $p$-value is $0.2070$ and $0.1212$, respectively.  It indicates that we 
cannot reject the null hypothesis at the 10\% level of significance that the data are coming from a GE distribution.

\section{\sc Optimality of Test Plan}
In the previous section we have obtained the Bayes estimates of the unknown parameters when the stress changing time $\tau_1$
is pre-fixed.   In this section we consider the problem of choosing the optimal value of $\tau_1$, for a simple step-stress 
experiment.  We obtain an optimal value of $\tau_1$ by minimizing the sum of posterior coefficient of 
variations of 
$\alpha,$ $\theta_1$ and $\theta_2$.  Since explicit form of the equation \eqref{eq:est1} cannot be obtained, we have used Lindley's approximation
to calculate the posterior coefficient of variations of the unknown parameters.  See Appendix \ref{th:lindly} for the detailed derivations of 
the Lindley's approximation.  
 By minimizing sum of the posterior coefficient of variations, an optimal value of $\tau_1$ can be obtained by using the following algorithm.

\begin{algo}
\item For given $\alpha,\theta_1, \theta_2, \tau_1$ and $n$ generate data from CEM.
\item Obtain the posterior variance of all the parameters using Lindley's approximation as explained in Appendix \ref{th:lindly}.
\item Repeat Step 1 and Step 2, $N$ times and take the average of variances.
\item Calculate the coefficients of variation for Bayes estimates of $\alpha,\theta_1, \theta_2. $\newline
Coefficient of Variation $= \frac{\text{posterior standard deviation}}{\text{posterior mean}} $
\item Take the sum of coefficients of variation for Bayes estimates of $\alpha, \theta_1$ and $\theta_2.$
\item Repeat Step 1 - Step 5 for different values of $\tau_1$ within its range.
\item Choose $\tau_1$ for which the sum of coefficients of variation is minimum.
\end{algo}

We have obtained numerically the optimal values of the stress changing times for different sample sizes and 
for different parameter values.  It has been observed that the posterior variance of $\alpha$ is decreasing with the increase
of $\tau_1$.  As expected the posterior variance of $\theta_1$ has a decreasing trend and the posterior variance
of $\theta_2$ increases with $\tau_1$.  However, if we consider total dispersion of three parameters in terms of coefficient
of variation, it is initially decreasing and then increasing as $\tau_1$ increases.  Hence, we have obtained a point where the total
dispersion is minimum and which is the optimal value of the stress changing time $\tau_1$.  The experimental results and the plots of the 
sum of the coefficient of variations are given below.

\begin{table}[!ht]\scriptsize
\caption{Optimal value of $\tau_1$ for different $n$ and $\alpha$ with $\theta_1=0.1,\theta_2=0.2$ }
\centering
\begin{tabular}{llc}
\toprule
$\alpha$ & n & Optimal value of $\tau_1$\\
\midrule
0.6 & 20 & 3.6 \\
0.6 & 30 & 6.4 \\
0.6 & 40 & 7.4 \\
0.6 & 50 & 7.2 \\
\\
1.0 & 20 & 8.4 \\
1.0 & 30 & 8.2 \\
1.0 & 40 & 9.4 \\
1.0 & 50 & 10.0 \\
\\
1.5 & 20 & 10.8 \\
1.5 & 30 & 13.0 \\
1.5 & 40 & 13.4 \\
1.5 & 50 & 13.4 \\
\bottomrule
\end{tabular}
\label{tab:optimtau}
\end{table}
\begin{figure}[h]
\begin{center}
 \subfigure[$\alpha=0.6, n=20$]{\includegraphics[width=1.5in,height=1in,angle=0]{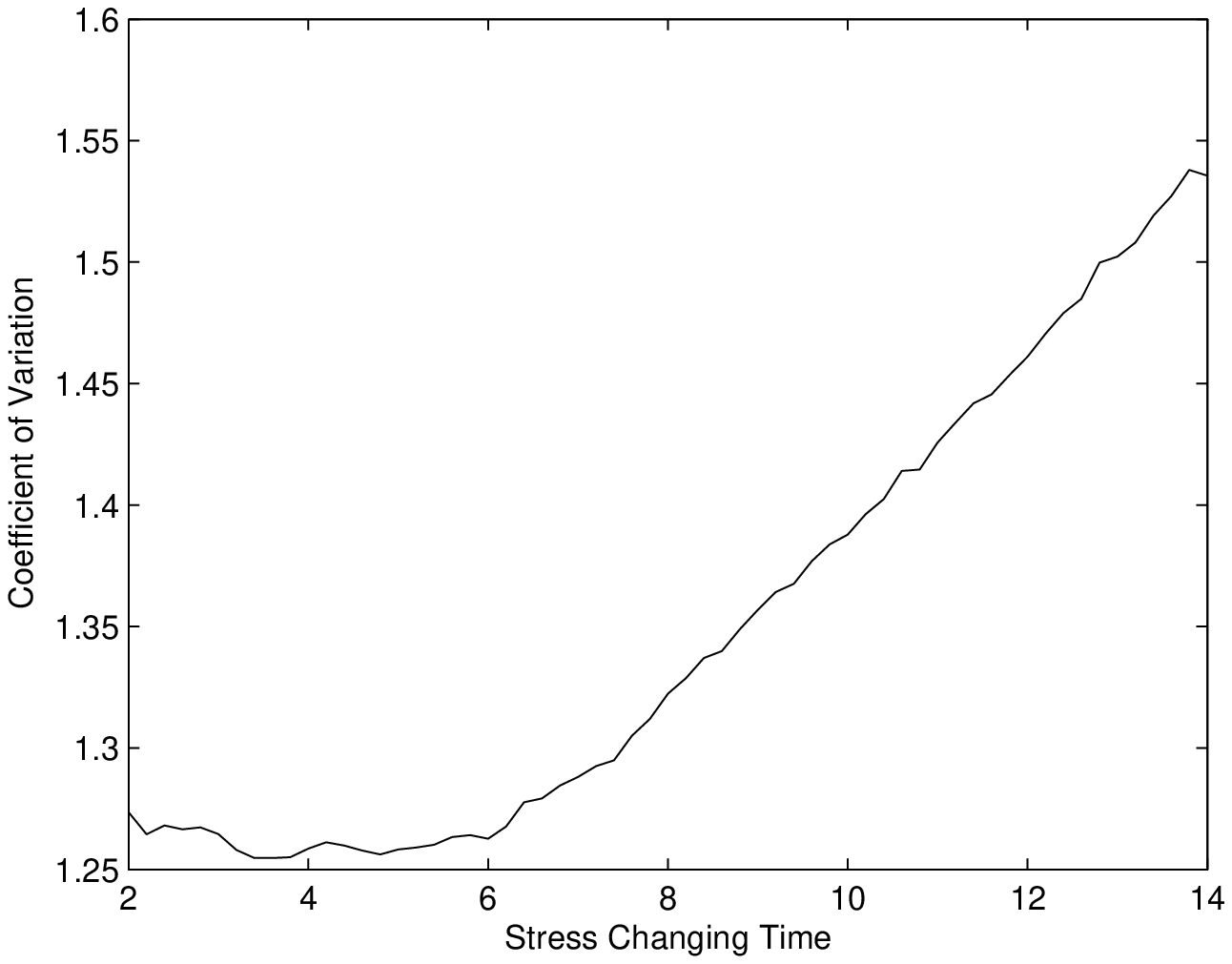}}
  \subfigure[$\alpha=0.6, n=30$]{\includegraphics[width=1.5in,height=1in,angle=0]{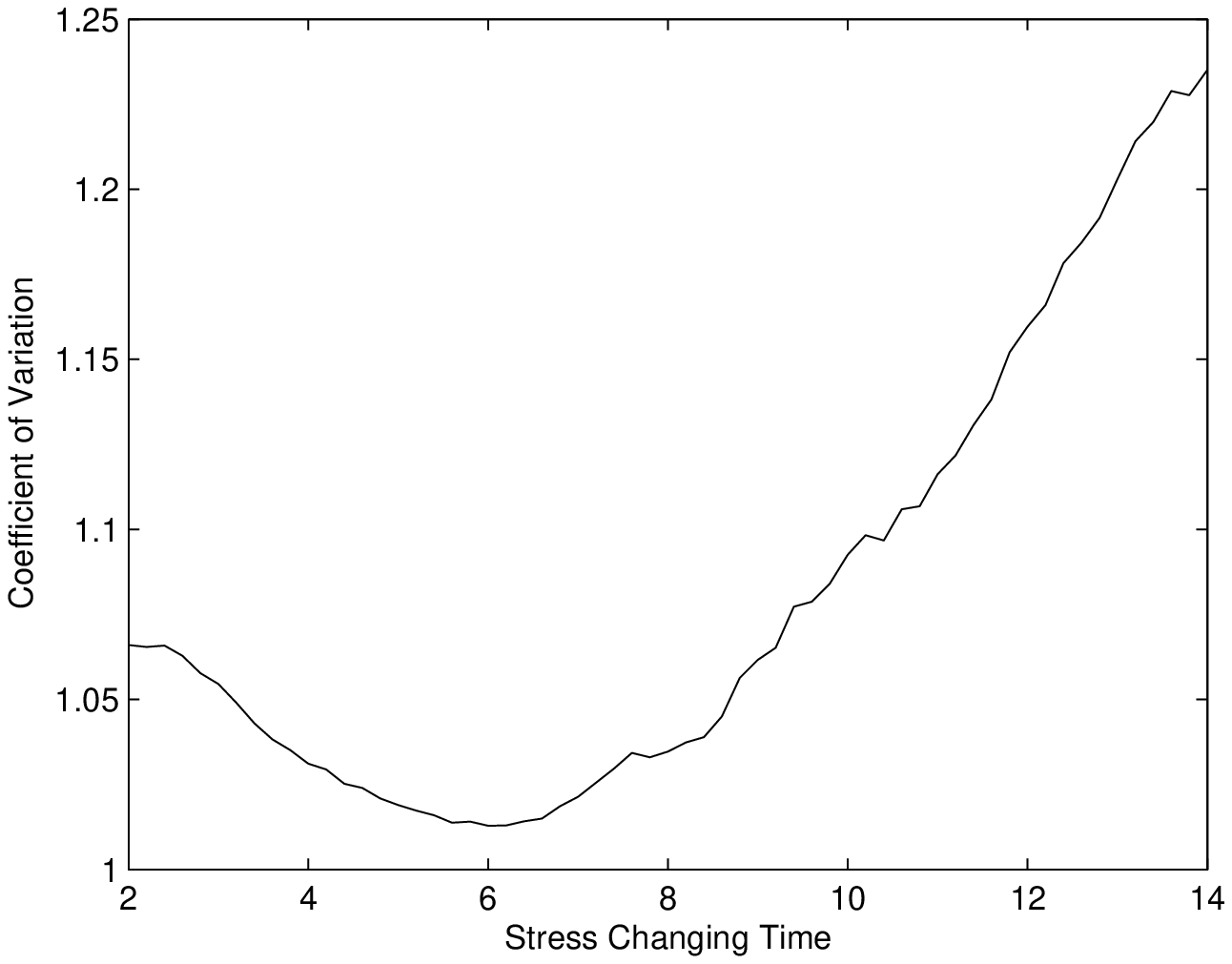}}
   \subfigure[$\alpha=0.6, n=40$]{\includegraphics[width=1.5in,height=1in,angle=0]{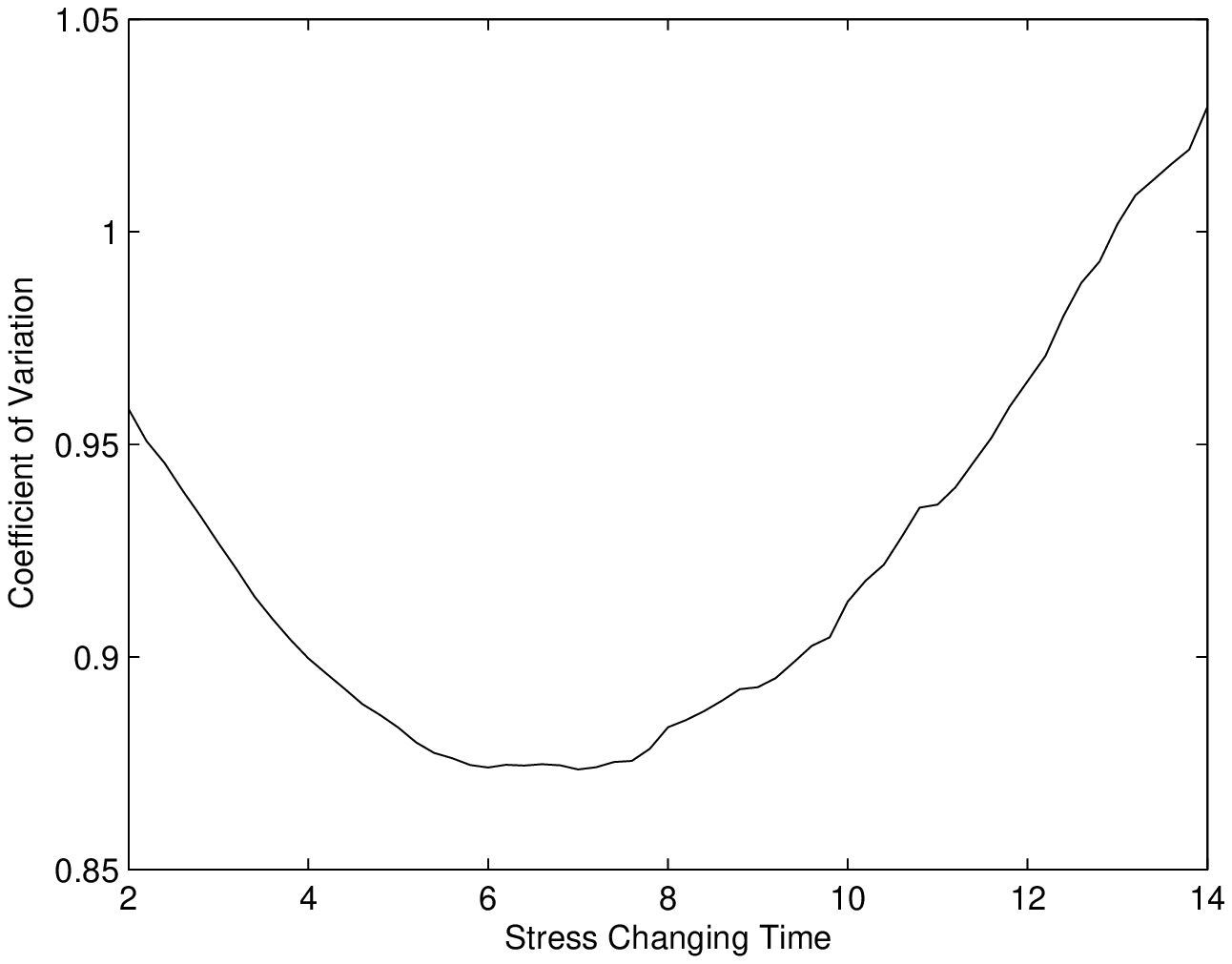}}
  \subfigure[$\alpha=0.6, n=50$]{\includegraphics[width=1.5in,height=1in,angle=0]{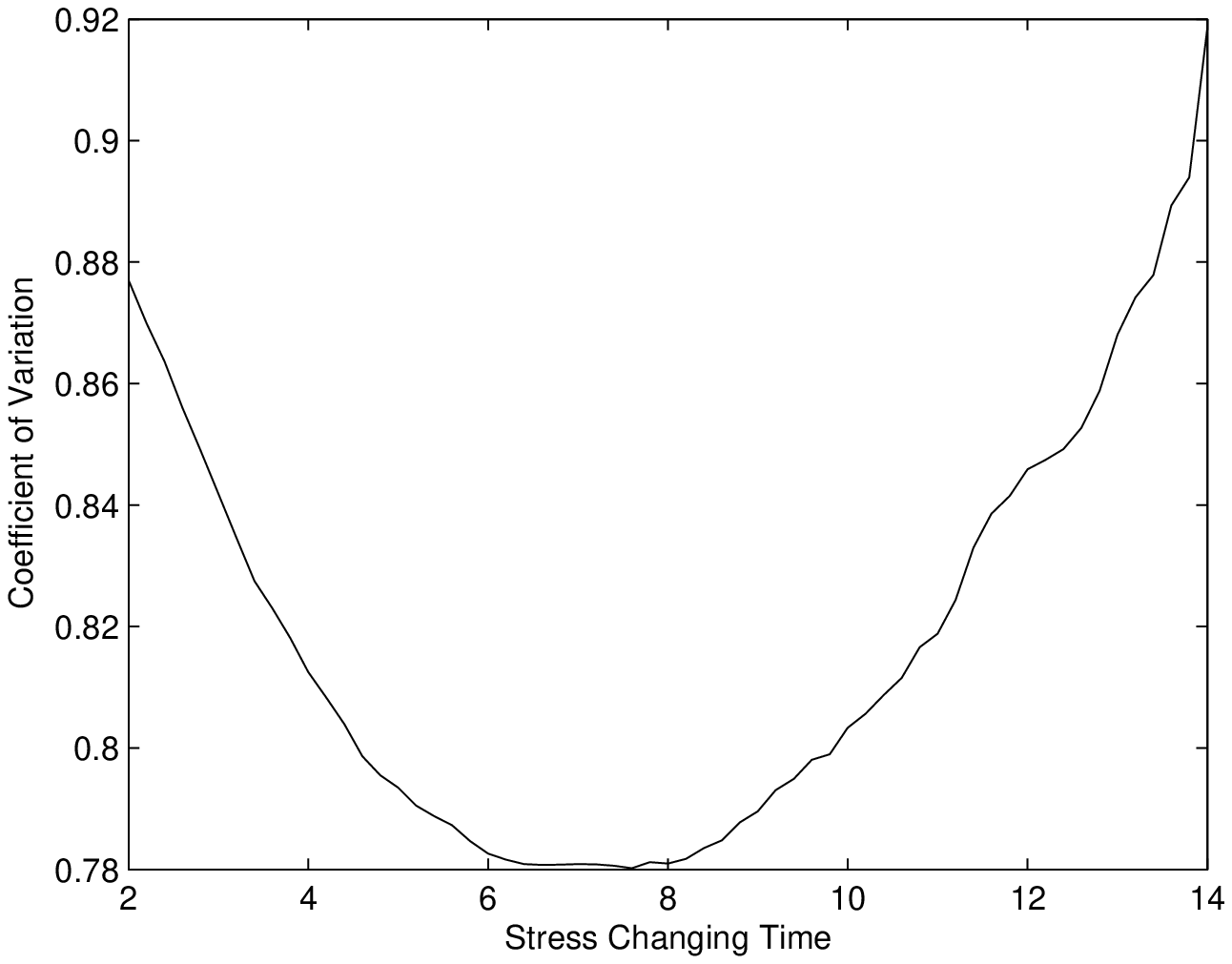}}
  \caption{Plot of total coefficient of variation for different values of  $\tau_1$ with parameter values $\alpha=0.6,\theta_1=0.1$ and $\theta_2=0.2$}
  \label{fig:optimality1}
\end{center}
\end{figure}

\begin{figure}[h]
\begin{center}
 \subfigure[$\alpha=1.0, n=20$]{\includegraphics[width=1.5in,height=1in,angle=0]{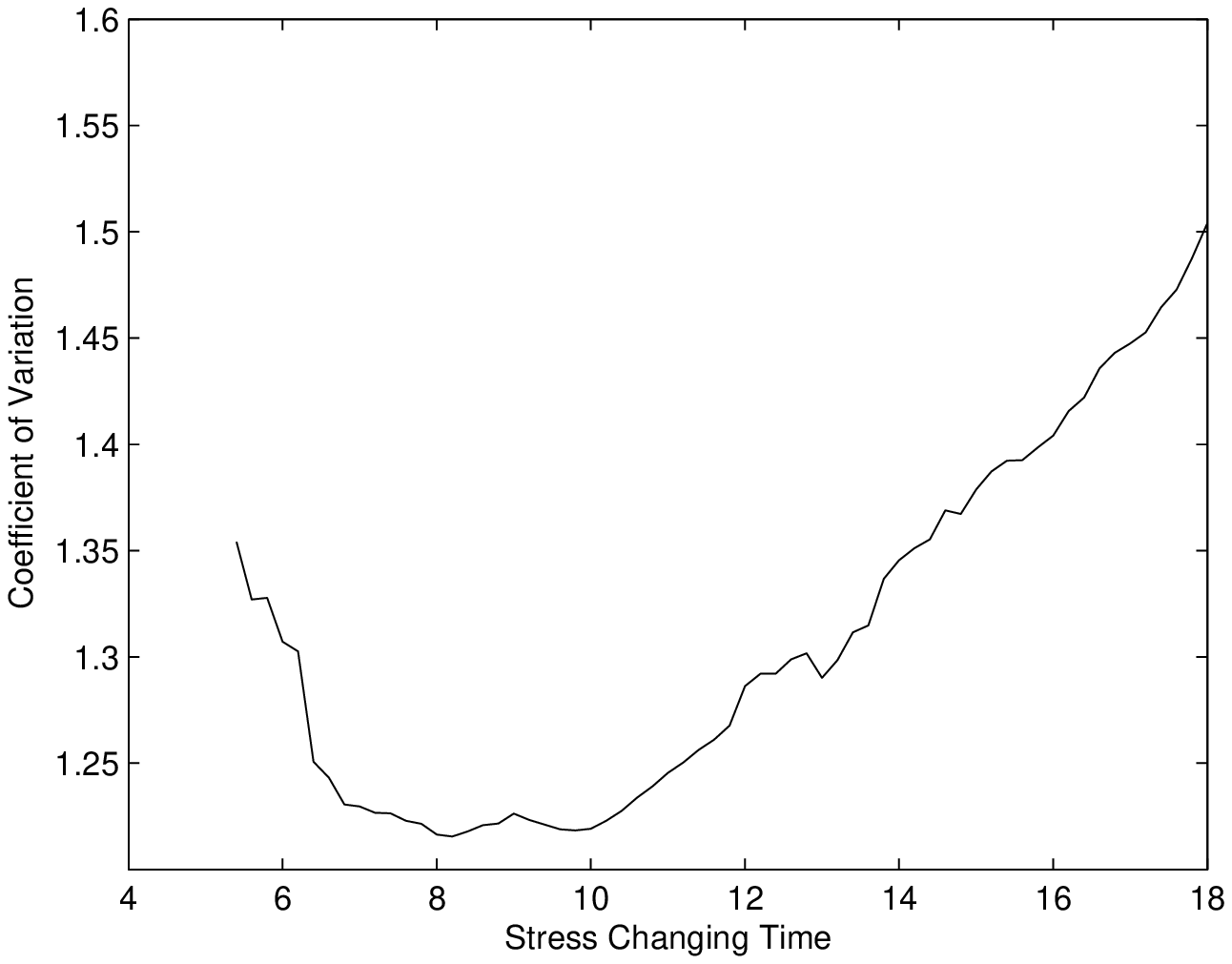}}
  \subfigure[$\alpha=1.0, n=30$]{\includegraphics[width=1.5in,height=1in,angle=0]{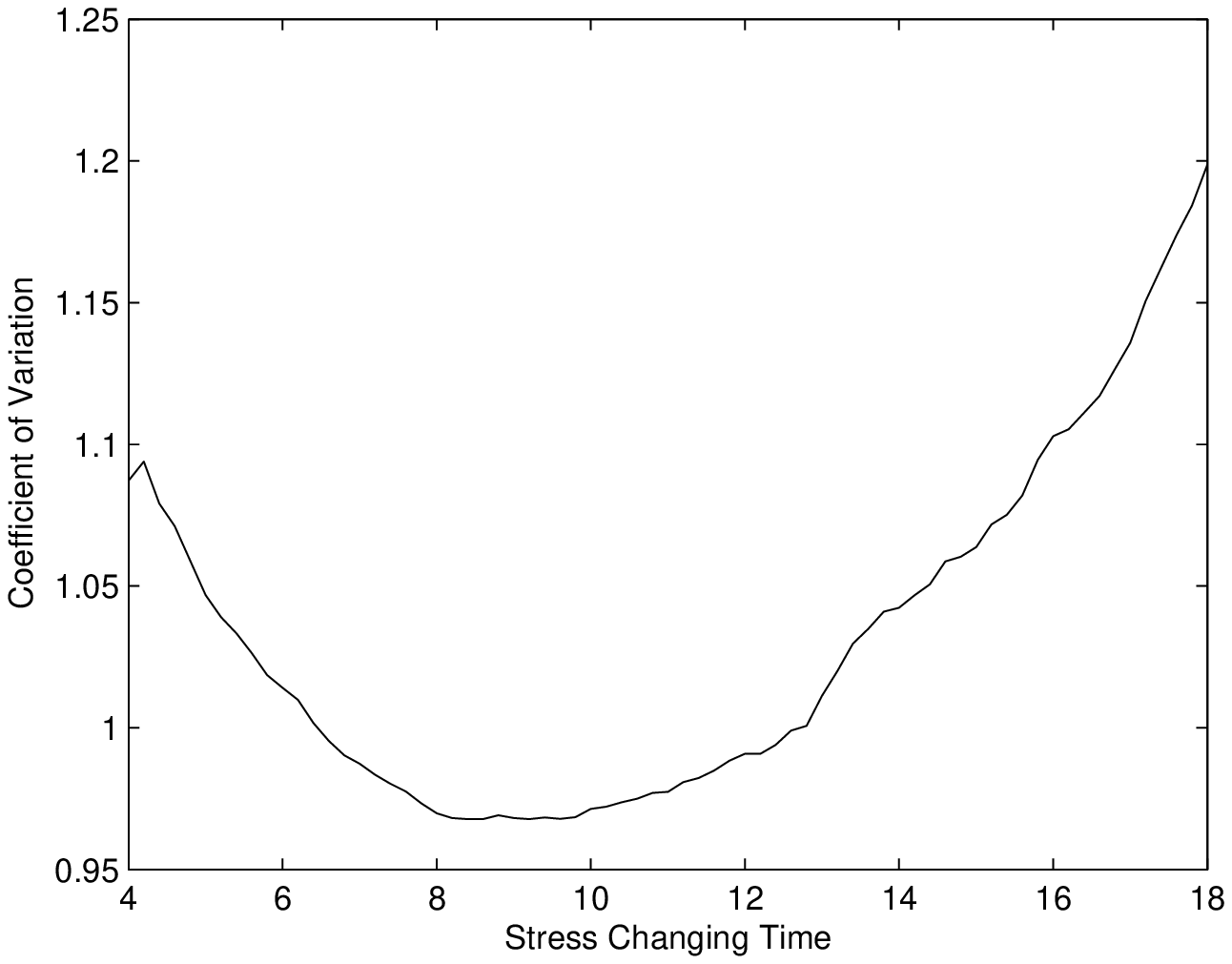}}
   \subfigure[$\alpha=1.0, n=40$]{\includegraphics[width=1.5in,height=1in,angle=0]{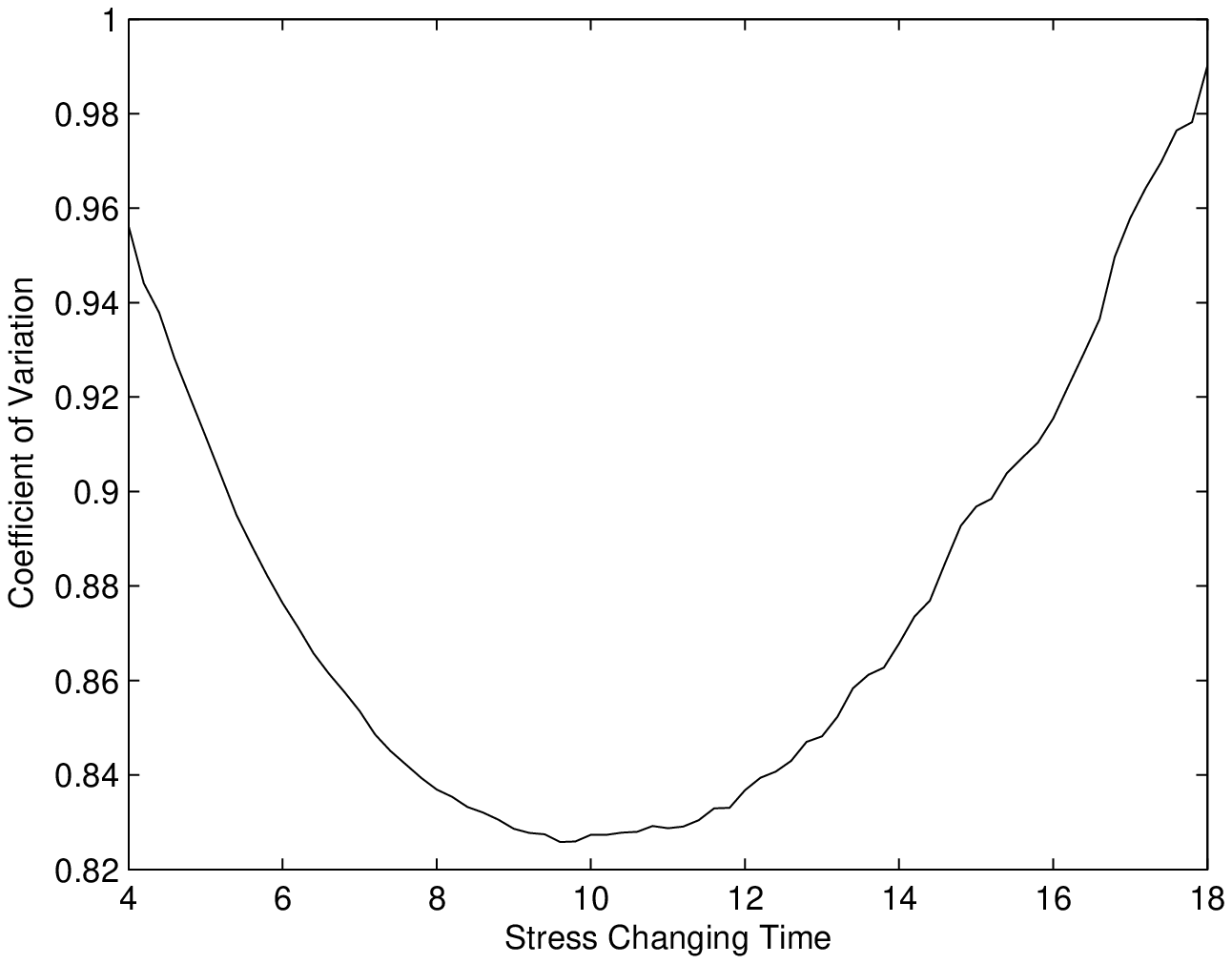}}
  \subfigure[$\alpha=1.0, n=50$]{\includegraphics[width=1.5in,height=1in,angle=0]{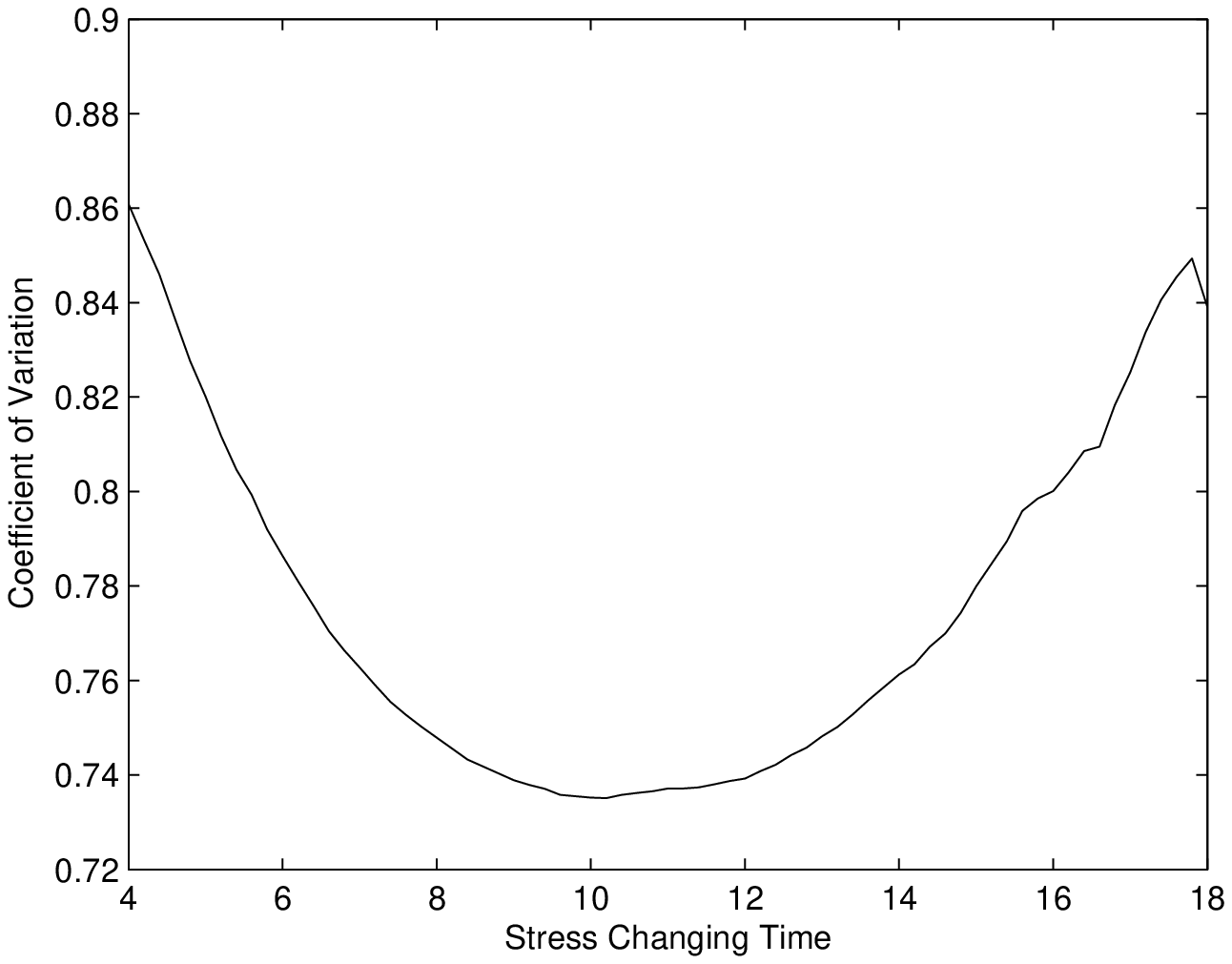}}
  \caption{Plot of total coefficient of variation for different values of  $\tau_1$ with parameter values $\alpha=1.0, 
\theta_1=0.1$ and $\theta_2=0.2$}
  \label{fig:optimality2}
\end{center}
\end{figure}

\begin{figure}[h]
\begin{center}
 \subfigure[$\alpha=1.5, n=20$]{\includegraphics[width=1.5in,height=1in,angle=0]{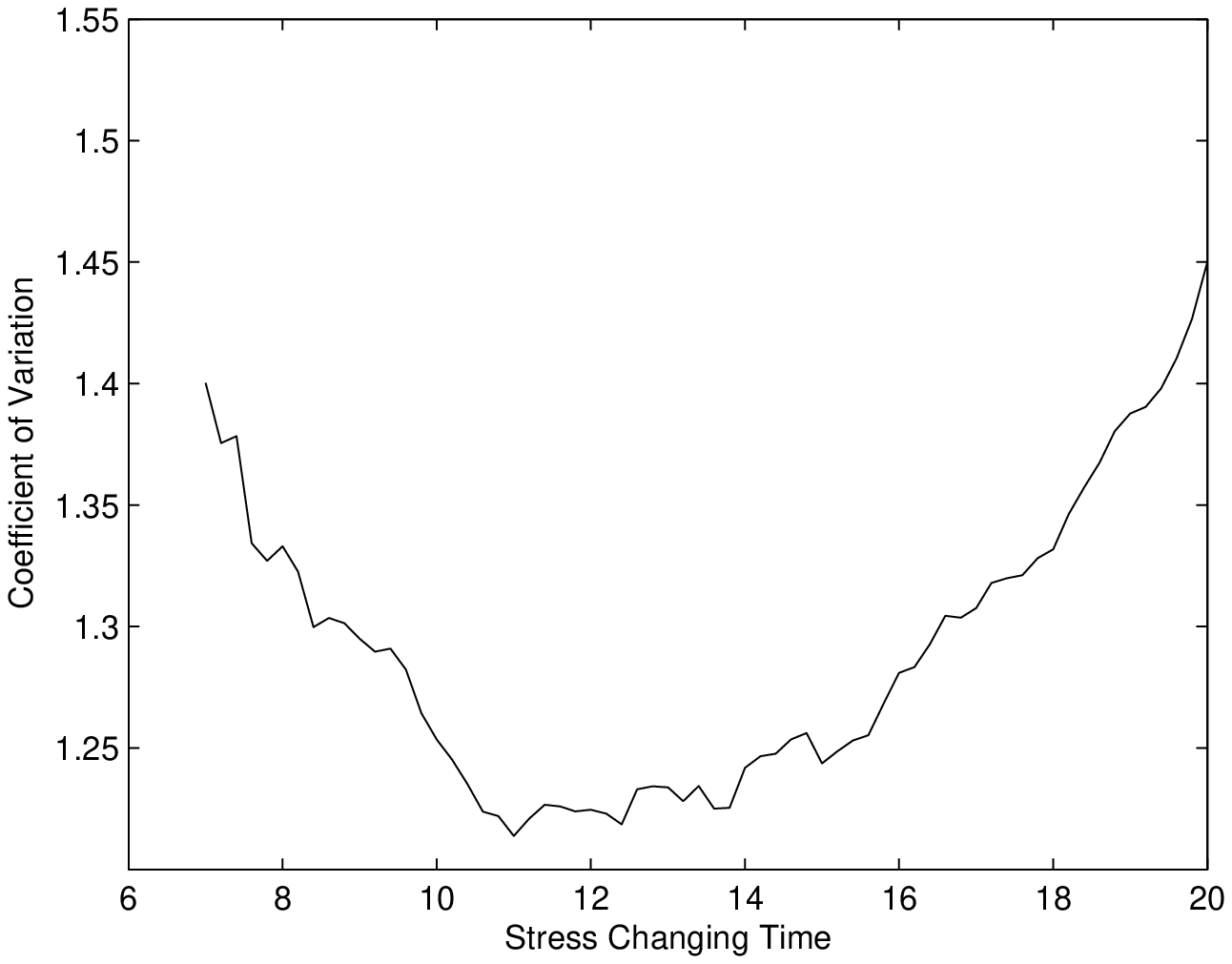}}
  \subfigure[$\alpha=1.5, n=30$]{\includegraphics[width=1.5in,height=1in,angle=0]{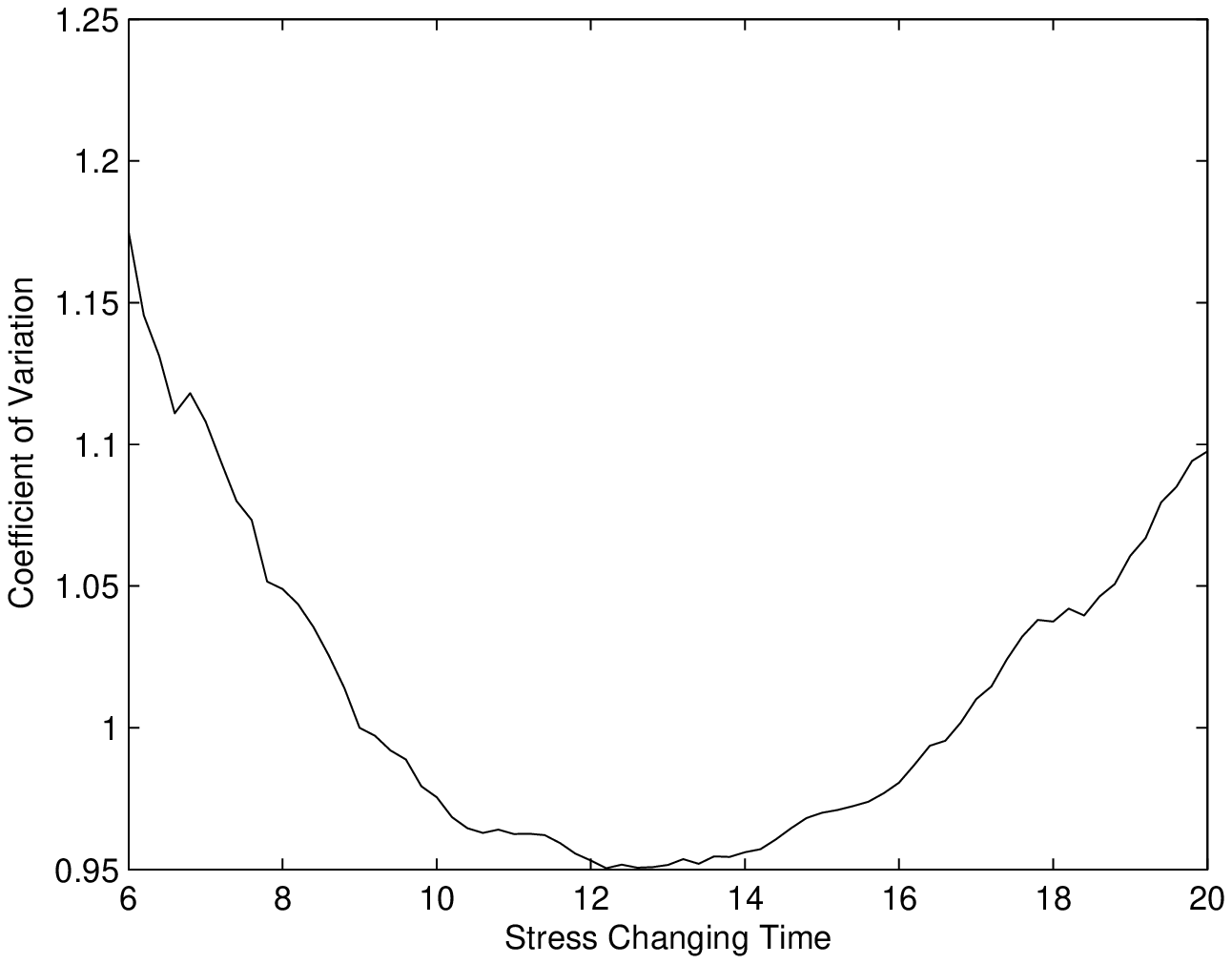}}
   \subfigure[$\alpha=1.5, n=40$]{\includegraphics[width=1.5in,height=1in,angle=0]{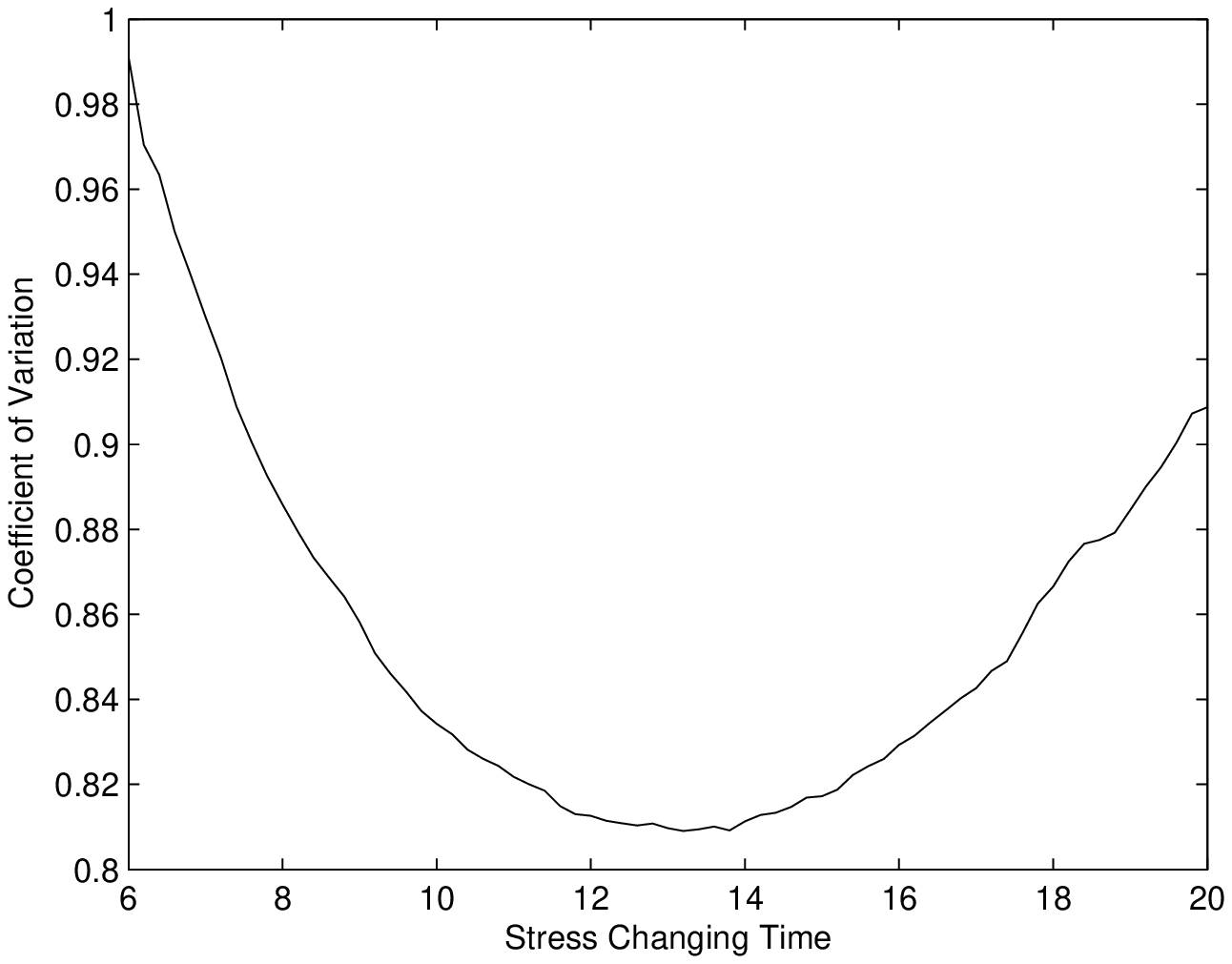}}
  \subfigure[$\alpha=1.5, n=50$]{\includegraphics[width=1.5in,height=1in,angle=0]{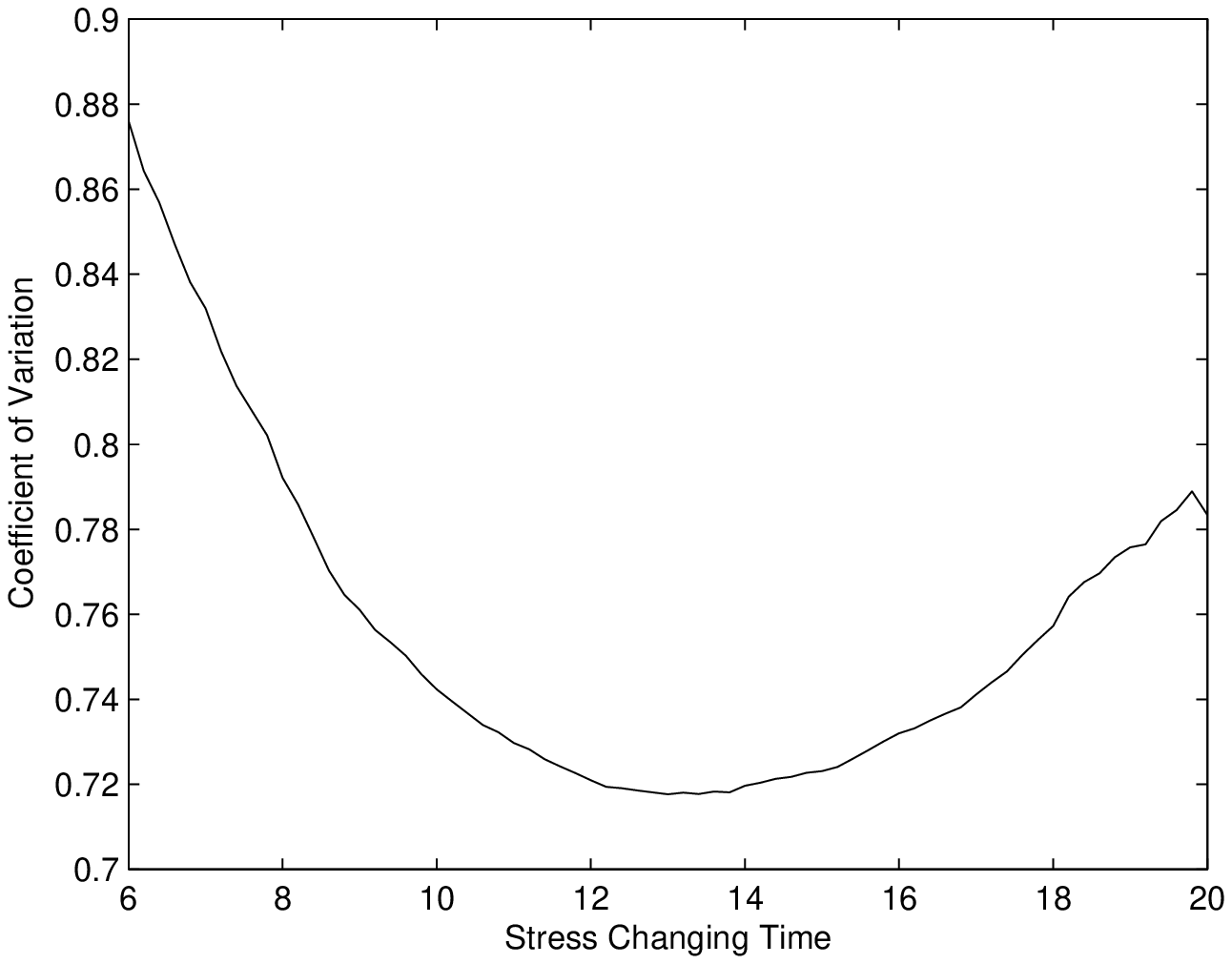}}
  \caption{Plot of total coefficient of variation for different values of  $\tau_1$ with parameter values $\alpha=1.5, 
\theta_1=0.1$ and $\theta_2=0.2$}
  \label{fig:optimality3}
\end{center}
\end{figure}

\section{\sc Conclusion}

In this paper we have considered the ordered restricted Bayesian inference of the unknown parameters of the GE distributions
when the data are coming from a step-step model.  It is assumed that the lifetime distribution satisfies the CEM assumptions. 
We have assumed a fairly flexible priors on the ordered parameters, and based on that we propose to use importance sampling
technique to compute the Bayes estimates and the associated credible intervals.  Extensive simulation experiments are performed
for different sample sizes and for different parametric values.  It is observed that the proposed method works quite well in 
practice.  Finally we consider choosing the optimal value for the stress changing time.  We choose the optimal value of 
optimal $\tau_1$, so that the sum of the posterior coefficient of variations is minimum.  Since the posterior coefficient of 
variations cannot be obtained in 
explicit forms, we suggest to use Lindley's approximation to compute the posterior coefficient of variations.  A small table is provided for 
optimal values of $\tau_1$, for different sample sizes and for different parametric values mainly for practical uses.

\section*{\sc Acknowledgements:} The authors would like thank two unknown reviewers for their valuable comments which have helped us 
to improve the manuscript significantly.

\bibliographystyle{plainnat-custom}
\bibliography{article,book}

\appendix
\section{\sc Appendix}
\subsection{\sc Three Parameters Lindley's Approximation }
\label{th:lindly}
For the three parameter case, using the notation $(\lambda_1,\lambda_2,\lambda_3) = (\alpha,\theta_2,\beta)$ Lindley's 
approximation of Bayes estimator of any function $g(\lambda_1,\lambda_2,\lambda_3)$ can be given by 
\begin{align}\label{eq:lind1}
\begin{array}{lll}
E_{\lambda_1,\lambda_2,\lambda_3\vert Data}\left(g(\lambda_1,\lambda_2,\lambda_3)\right)
& = & g(\hat{\lambda}_1,\hat{\lambda}_2,\hat{\lambda}_3) + \frac{1}{2}\sum_{i=1}^{3}\sum_{j=1}^{3} u_{ij}\sigma_{ij} 
+ \sum_{i=1}^{3}\sum_{j=1}^{3} u_{i}\rho_{j}\sigma_{ij}  \\ 
{} & {} & + \frac{1}{2} \sum_{i=1}^{3}\sum_{j=1}^{3} \sum_{k=1}^{3} L_{ijk}U_{k}\sigma_{ij}, 
\end{array}
\end{align}
\begin{align}
\shortintertext{where}
 L_{ijk} & = \frac{\delta^3L}{\delta\lambda_i\delta\lambda_j\delta\lambda_k} ; \quad\quad i,j,k = 1(1)3 \quad \text{and L is log likelihood of the data;} \nonumber  \\
 u_i & = \frac{\delta g}{\delta \lambda_i}; \quad \quad i = 1(1)3; \nonumber \\
 u_{ij} & = \frac{\delta^2 g}{\delta \lambda_i \delta \lambda_j}; \quad \quad i,j = 1(1)3; \nonumber \\
 \sigma_{ij} & = (i,j)^{th} \text{ element of the inverse of the matrix having elements} ((-L_{ij})); \nonumber \\ 
 \rho_i & = \frac{\delta log \pi}{\delta \lambda_i}; \nonumber \\
 U_k & = \sum_{i=1}^{3} u_i \sigma_{ki} ; \quad \quad k=1(1)3, \nonumber
\end{align}
$\hat{\lambda}_1,\hat{\lambda}_2,\hat{\lambda}_3$  are MLEs of $\lambda_1,\lambda_2,\lambda_3$ respectively and all of the quantities are evaluated
at $(\hat{\lambda}_1,\hat{\lambda}_2,\hat{\lambda}_3).$ The log likelihood function of the data under order restriction of parameter is given by
\begin{align}\label{eq:lind2}
 L(t_1,t_2,\ldots t_n \vert \alpha, \theta_2, \beta) = & log(n!) + n log(\alpha) + n_1 log(\beta) + n log(\theta_2) - \beta \theta_2 \sum_{k=1}^{n_1}t_k \nonumber \\
 {} & + (\alpha-1) \sum_{k=1}^{n_1}log(1-e^{-\beta\theta_2 t_k})- \theta_2\sum_{k=n_1+1}^{n}(t_k+\beta\tau_1-\tau_1) \nonumber \\
 {} & +(\alpha-1)  \sum_{k=n_1+1}^{n}log(1-e^{-\theta_2(t_k+\beta\tau_1-\tau_1)}).
\end{align}
The MLEs of $\alpha, \theta_2,$ and $\beta$ can be obtain by maximizing (\ref{eq:lind2}).
\begin{align}\label{eq:order1}
\frac{\delta L}{\delta \alpha} = 0 \Rightarrow 
 \hat{\alpha} = \frac{-n}{\sum_{k=1}^{N_1}ln(1-e^{-\beta\theta_2 t_k})+\sum_{k=N_1+1}^{n}ln(1-e^{-\theta_2(t_k+\beta\tau_1-\tau_1)})},
\end{align}
\begin{align}\label{eq:order2}
\frac{\delta L}{\delta \theta_2} = 0 \Rightarrow \nonumber
\frac{n}{\theta_2} - \sum_{k=1}^{N_1}\beta t_k + (\alpha-1)\sum_{k=1}^{N_1}\frac{\beta t_k e^{-\beta \theta_2 t_k}}{1-e^{-\beta \theta_2 t_k}} \nonumber
-\sum_{k=N_1+1}^{n}(t_k+\beta \tau_1 - \tau_1) \\ 
+(\alpha-1)\sum_{k=N_1+1}^{n}\frac{(t_k+\beta \tau_1 - \tau_1)e^{-\theta_2(t_k+\beta\tau_1-\tau_1)}}{1-e^{-\theta_2(t_k+\beta \tau_1-\tau_1)}} = 0.
\end{align}
For known $\beta(0<\beta<1),$ the estimate of $\theta_2$ can be obtain by solving \eqref{eq:order2} numerically and hence an estimate of $\alpha$
from \eqref{eq:order1}. The value of $\beta$ between $0$ and $1$ and the corresponding estimates of $\alpha$ and $\theta_2,$ for which
likelihood is maximum will be the MLEs of $\beta, \alpha$ and $\theta_2$ respectively.

\begin{eqnarray}
L_{11} & =  & \frac{\delta^2{l}}{\delta \alpha^2} = -\frac{n}{\alpha^2}, \nonumber  \\
L_{12} & = & L_{21} = \frac{\delta^2{l}}{\delta \alpha \delta \theta_2} = \sum_{k=1}^{n_1}\frac{\beta t_k e^{-\beta \theta_2 t_k}}{1-e^{-\beta \theta_2 t_k}} 
+\sum_{k=n_1+1}^{n}\frac{(t_k+\beta\tau_1-\tau_1) e^{-\theta_2(t_k+\beta\tau_1-\tau_1)}}{1-e^{-\theta_2(t_k+\beta\tau_1-\tau_1)}}, \nonumber  \\
L_{13}& = & L_{31} = \frac{\delta^2{l}}{\delta \alpha \delta \beta} = \sum_{k=1}^{n_1}\frac{\theta_2 t_k e^{-\beta \theta_2 t_k}}{1-e^{-\beta \theta_2 t_k}} 
+ \sum_{k=n_1+1}^{n}\frac{\theta_2\tau_1 e^{-\theta_2(t_k+\beta\tau_1-\tau_1)}}{1-e^{-\theta_2(t_k+\beta\tau_1-\tau_1)}}, \nonumber  \\
L_{22} & = & \frac{\delta^2{l}}{\delta \theta_2^2} = -\frac{n}{\theta_2^2}-(\alpha-1)\sum_{k=1}^{n_1}\frac{\theta_2^2 t_k^2 e^{-\beta \theta_2 t_k}}{(1-e^{-\beta \theta_2 t_k})^2} 
-(\alpha-1) \sum_{k=n_1+1}^{n}\frac{(t_k+\beta\tau_1-\tau_1)^2 e^{-\theta_2(t_k+\beta\tau_1-\tau_1)}}{(1-e^{-\theta_2(t_k+\beta\tau_1-\tau_1)})^2}, \nonumber \\
L_{23} & = & L_{32}=\frac{\delta^2{l}}{\delta \theta_2 \delta \beta} = -\alpha \sum_{k=1}^{n_1}t_k-(\alpha-1)\sum_{k=1}^{n_1}\frac{t_k (1-e^{-\beta \theta_2 t_k})-\beta \theta_2 t_k^2 e^{-\beta \theta_2 t_k}}{(1-e^{-\beta \theta_2 t_k})^2}
-\alpha(n-n_1)\tau_1  \nonumber \\
&  & -(\alpha-1) \sum_{k=n_1+1}^{n} 
\frac{\tau_1(1- e^{-\theta_2(t_k+\beta\tau_1-\tau_1)})-\theta_2 \tau_1(t_k+\beta\tau_1-\tau_1) 
e^{-\theta_2(t_k+\beta\tau_1-\tau_1)} }
{(1-e^{-\theta_2(t_k+\beta\tau_1-\tau_1)})^2}, \nonumber \\
L_{33} & = & \frac{\delta^2{l}}{\delta \beta^2} = -\frac{n_1}{\beta^2}-(\alpha-1)\sum_{k=1}^{n_1}\frac{\theta_2^2 t_k^2 e^{-\beta \theta_2 t_k}}{(1-e^{-\beta \theta_2 t_k})^2} 
-(\alpha-1) \sum_{k=n_1+1}^{n}\frac{\theta_2^2 \tau_1^2 e^{-\theta_2(t_k+\beta\tau_1-\tau_1)}}{(1-e^{-\theta_2(t_k+\beta\tau_1-\tau_1)})^2}, \nonumber  
\end{eqnarray}
\begin{eqnarray}
L_{111} & = & \frac{\delta^3L}{\delta\alpha^3} = \frac{2n}{\alpha^3}, \quad L_{112} =\frac{\delta^3L}{\delta\alpha^2\delta\theta_2}= 0, \quad L_{113} =\frac{\delta^3L}{\delta\alpha^2\delta\beta}= 0, \nonumber  \\
L_{122} & = & \frac{\delta^3L}{\delta\alpha\delta\theta_2^2} = -\sum_{k=1}^{n_1}\frac{\beta^2 t_k^2 e^{-\beta \theta_2 t_k}}{(1-e^{-\beta \theta_2 t_k})^2} - \sum_{k=n_1+1}^{n}\frac{(t_k+\beta\tau_1-\tau_1)^2 e^{-\theta_2(t_k+\beta\tau_1-\tau_1)}}{(1-e^{-\theta_2(t_k+\beta\tau_1-\tau_1)})^2},   \nonumber \\  
L_{123} & = & \frac{\delta^3L}{\delta\alpha\delta\theta_2\delta\beta}  = \sum_{k=1}^{n_1}[\frac{t_k(1-e^{-\beta \theta_2 t_k})-\theta_2\beta t_k^2e^{-\beta\theta_2 t_k}}{(1-e^{-\beta \theta_2 t_k})^2} - t_k ]   \nonumber \\
&  &  +\sum_{k=n_1+1}^{n}[\frac{\tau_1(1-e^{-\theta_2(t_k+\beta\tau_1-\tau_1)})-\theta_2 \tau_1(t_k+\beta\tau_1-\tau_1) e^{-\theta_2(t_k+\beta\tau_1-\tau_1)}}{(1-e^{-\theta_2(t_k+\beta\tau_1-\tau_1)})^2} - \tau_1],  \nonumber  \\
L_{133} & = & \frac{\delta^3L}{\delta\alpha\delta\beta^2} = -\sum_{k=1}^{n_1}\frac{\theta_2^2 t_k^2 e^{-\beta \theta_2 t_k}}{(1-e^{-\beta \theta_2 t_k})^2} - \sum_{k=n_1+1}^{n}\frac{\theta_2^2 \tau_1^2 e^{-\theta_2(t_k+\beta\tau_1-\tau_1)}}{(1-e^{-\theta_2(t_k+\beta\tau_1-\tau_1)})^2},  \nonumber \\
 L_{222} & = & \frac{\delta^3L}{\delta\theta_2^3}  = \frac{2n}{\theta_2^3} + (\alpha-1)\sum_{k=1}^{n_1}\frac{\beta^3 t_k^3 e^{-\beta \theta_2 t_k}(1+e^{-\beta \theta_2 t_k})}{(1-e^{-\beta \theta_2 t_k})^3} \nonumber \\ 
&  & +(\alpha-1)\sum_{k=n_1+1}^{n}\frac{(t_k+\beta\tau_1-\tau_1)^3 e^{-\theta_2(t_k+\beta\tau_1-\tau_1)}(1+e^{-\theta_2(t_k+\beta\tau_1-\tau_1)})}{(1-e^{-\theta_2(t_k+\beta\tau_1-\tau_1)})^3},   \nonumber  \\
 L_{223} & = & \frac{\delta^3L}{\delta\theta_2^2\delta\beta}  = -(\alpha-1)\sum_{k=1}^{n_1}\frac{2\beta t_k^2 e^{-\beta \theta_2 t_k}}{(1-e^{-\beta \theta_2 t_k})^2}+(\alpha-1)\sum_{k=1}^{n_1}\frac{\beta^2 \theta_2 t_k^3 e^{-\beta \theta_2 t_k}(1+e^{-\beta \theta_2 t_k})}{(1-e^{-\beta \theta_2 t_k})^3} \nonumber \\
&  & -(\alpha-1)\sum_{k=n_1+1}^{n}\frac{2\tau_1(t_k+\beta\tau_1-\tau_1) e^{-\theta_2(t_k+\beta\tau_1-\tau_1)}}{(1-e^{-\theta_2(t_k+\beta\tau_1-\tau_1)})^2}  \nonumber  \\
&  & +(\alpha-1)\sum_{k=n_1+1}^{n}\frac{\theta_2 \tau_1 (t_k+\beta\tau_1-\tau_1)^2 e^{-\theta_2(t_k+\beta\tau_1-\tau_1)}(1+e^{-\theta_2(t_k+\beta\tau_1-\tau_1)})}{(1-e^{-\theta_2(t_k+\beta\tau_1-\tau_1)})^3},   \nonumber    
\end{eqnarray}
\begin{eqnarray}
L_{233} & = & \frac{\delta^3L}{\delta\theta_2\delta\beta^2}  = -(\alpha-1)\sum_{k=1}^{n_1}\frac{2\theta_2 t_k^2 e^{-\beta \theta_2 t_k}}{(1-e^{-\beta \theta_2 t_k})^3} +(\alpha-1)\sum_{k=1}^{n_1}\frac{\beta \theta_2^2 t_k^3 e^{-\beta \theta_2 t_k}(1+e^{-\beta \theta_2 t_k})}{(1-e^{-\beta \theta_2 t_k})^3} \nonumber \\ 
&  & -(\alpha-1)\sum_{k=n_1+1}^{n}\frac{2\theta_2\tau_1^2 e^{-\theta_2(t_k+\beta\tau_1-\tau_1)}}{(1-e^{-\theta_2(t_k+\beta\tau_1-\tau_1)})^3}  \nonumber  \\
&  & +(\alpha-1)\sum_{k=n_1+1}^{n}\frac{\theta_2^2 \tau_1^2 (t_k+\beta\tau_1-\tau_1) e^{-\theta_2(t_k+\beta\tau_1-\tau_1)}(1+e^{-\theta_2(t_k+\beta\tau_1-\tau_1)})}{(1-e^{-\theta_2(t_k+\beta\tau_1-\tau_1)})^3},  \nonumber  \\  
 L_{333} & = & \frac{\delta^3L}{\delta\beta^3}  =\frac{2n_1}{\beta^3} +(\alpha-1)\sum_{k=1}^{n_1}\frac{\theta_2^3 t_k^3 e^{-\beta \theta_2 t_k}(1+e^{-\beta \theta_2 t_k})}{(1-e^{-\beta \theta_2 t_k})^3} \nonumber \\
&  & +(\alpha-1)\sum_{k=n_1+1}^{n}\frac{\theta_2^3 \tau_1^3 e^{-\theta_2(t_k+\beta\tau_1-\tau_1)}(1+e^{-\theta_2(t_k+\beta\tau_1-\tau_1)})}{(1-e^{-\theta_2(t_k+\beta\tau_1-\tau_1)})^3}.  \nonumber
\end{eqnarray}
Note that $L_{ijk}$ does not depends on the order of appearance of $i, j$ and $k.$
\begin{align}
 \rho_1 = \frac{\delta log(\pi)}{\delta \alpha} = \frac{b_0-1}{\alpha} - a_0; \quad \rho_2 = \frac{\delta log(\pi)}{\delta \theta_2} = \frac{b_1-1}{\theta_2} - a_1 \quad \text{and} \quad \rho_3 = \frac{\delta log(\pi)}{\delta \beta} = \frac{a_2-1}{\beta} - \frac{b_2-1}{1-\beta}  \nonumber 
\end{align}

To obtain the posterior variance of the parameters we need to take below assumptions on the function $g(\alpha,\theta_2,\beta).$ \newline
$(a)$ \quad To calculate posterior variance of $\alpha :$ \quad $g = \alpha$ and $g=\alpha^2.$  \newline
$(b)$ \quad To calculate posterior variance of $\theta_1 :$ \quad $g = \beta \theta_2$ and $g=\beta^2 \theta_2^2.$  \newline
$(c)$ \quad To calculate posterior variance of $\theta_2 :$ \quad $g = \theta_2$ and $g=\theta_2^2.$  \newline
In case $(a)$ if $g=\alpha,$ \quad $u_1 = 1, \quad u_2=u_3=0, \quad u_{ij} = 0, i,j=1(1)3$. \newline
If $g=\alpha^2,$ \quad $u_1 = 2\alpha, \quad u_2=u_3=0, \quad u_{11}=2, \quad u_{ij}=0$ for $i,j=1(1)3$ and $(i,j)\neq(1,1).$\newline
In case $(b)$ if $g=\beta \theta_2,$ \quad $u_1 = 0$, $u_2 = \beta$, $u_3=\theta_2,  \quad u_{23}=1, u_{ij} = 0,$ for  $i,j=1(1)3$ and 
$(i,j)\neq(2,3).$ \newline
If $g=\beta^2 \theta_2^2,$ \quad $u_1=0$, $u_2 = 2\beta^2\theta_2$, $u_3 = 2\beta\theta_2^2, \quad u_{11}=u_{12}=u_{13}=0, u_{22}=2\beta^2, u_{23} = 4\beta\theta_2, u_{33}=2\theta_2^2.$  \newline
In case $(c)$ if $g=\theta_2,$ \quad $u_2=1, u_1=u_3=0 \quad  u_{ij} = 0, i,j=1(1)3.$\newline
If $g=\theta_2^2,$ \quad $u_2 = 2\theta_2, \quad u_1=u_3=0, \quad u_{22}=2, \quad u_{ij}=0$ for $i,j=1(1)3$ and $(i,j)\neq(2,2).$\newline
Note that $u_{ij} = u_{ji}$ for all $i,j = 1(1)3.$ \newline
Now posterior variance of the parameters can be obtain by using the equation (\ref{eq:lind1}).

\newpage

\subsection{\sc Simulation Results}
\label{th:simulation}
\begin{table}[!ht]\scriptsize
\caption{AEs and MSEs of $\alpha$, $\theta_1$, and $\theta_2$ based on $5000$ simulations with $\alpha=1.5$,
$\theta_1=0.1$, and $\theta_2=0.2$ for different values of $n$, $\tau_1$ and $\tau_2$ of Type-I censored data.}
\centering
\begin{tabular}{*{9}{c}}
\toprule
\multicolumn{3}{c}{} & \multicolumn{2}{c}{$\alpha$} & \multicolumn{2}{c}{$\theta_1$} &
\multicolumn{2}{c}{$\theta_2$}\\
\cmidrule(lr){4-5}\cmidrule(lr){6-7}\cmidrule(lr){8-9}
\multicolumn{1}{c}{$n$} & \multicolumn{1}{c}{$\tau_1$} & \multicolumn{1}{c}{$\tau_2$} & \multicolumn{1}{c}{AE} & \multicolumn{1}{c}{MSE} &
\multicolumn{1}{c}{AE} & \multicolumn{1}{c}{MSE} & \multicolumn{1}{c}{AE} & \multicolumn{1}{c}{MSE} \\
\midrule
20	&	     7	&	    13	&	1.8329	&	0.8966	&	0.1114	&	0.0018	&	0.2079	&	0.0043  \\
{}	&	     9	&	    13	&	1.8214	&	0.8455	&	0.1096	&	0.0016	&	0.2140	&	0.0062  \\
{}	&	     9	&	    15	&	1.8132	&	0.7975	&	0.1080	&	0.0015	&	0.2103	&	0.0051  \\[3mm]
30	&	     7	&	    13	&	1.7538	&	0.4748	&	0.1111	&	0.0014	&	0.2052	&	0.0029  \\
{}	&	     9	&	    13	&	1.7314	&	0.4446	&	0.1088	&	0.0011	&	0.2067	&	0.0038  \\
{}	&	     9	&	    15	&	1.6914	&	0.3702	&	0.1071	&	0.0010	&	0.2059	&	0.0030  \\[3mm]
40	&	     7	&	    13	&	1.7024	&	0.3247	&	0.1106	&	0.0011	&	0.2040	&	0.0020  \\
{}	&	     9	&	    13	&	1.6671	&	0.2831	&	0.1078	&	0.0008	&	0.2043	&	0.0028  \\
{}	&	     9	&	    15	&	1.6541	&	0.2562	&	0.1064	&	0.0008	&	0.2004	&	0.0021  \\[3mm]
50	&	     7	&	    13	&	1.6716	&	0.2261	&	0.1105	&	0.0009	&	0.2017	&	0.0016  \\
{}	&	     9	&	    13	&	1.6693	&	0.2228	&	0.1099	&	0.0007	&	0.2029	&	0.0022  \\
{}	&	     9	&	    15	&	1.6395	&	0.1909	&	0.1075	&	0.0007	&	0.2011	&	0.0019  \\
\bottomrule
\end{tabular}
\label{tab:AE4}
\end{table}

\begin{table}[!ht]\scriptsize
\caption{AEs and MSEs of $\alpha$, $\theta_1$, and $\theta_2$ based on $5000$ simulations with $\alpha=1.5$,
$\theta_1=0.1$, and $\theta_2=0.2$ for different values of $n$, $\tau_1$ and $r$ of Type-II censored data.}
\centering
\begin{tabular}{*{9}{c}}
\toprule
\multicolumn{3}{c}{} & \multicolumn{2}{c}{$\alpha$} & \multicolumn{2}{c}{$\theta_1$} &
\multicolumn{2}{c}{$\theta_2$}\\
\cmidrule(lr){4-5}\cmidrule(lr){6-7}\cmidrule(lr){8-9}
\multicolumn{1}{c}{$n$} & \multicolumn{1}{c}{$\tau_1$} & \multicolumn{1}{c}{$r$} & \multicolumn{1}{c}{AE} & \multicolumn{1}{c}{MSE} &
\multicolumn{1}{c}{AE} & \multicolumn{1}{c}{MSE} & \multicolumn{1}{c}{AE} & \multicolumn{1}{c}{MSE} \\
\midrule
20	&	     7	&	    15	&	1.9245	&	1.2123	&	0.1148	&	0.0022	&	0.2273	&	0.0107  \\
{}      &	     9	&	    15	&	1.8758	&	0.9971	&	0.1115	&	0.0017	&	0.2406	&	0.0400  \\
{}      &	     9	&	    17	&	1.8348	&	0.8609	&	0.1104	&	0.0017	&	0.2227	&	0.0081  \\[3mm]
30  	&	     7	&	    23	&	1.7902	&	0.5586	&	0.1133	&	0.0016	&	0.2142	&	0.0040  \\
{}      &	     9	&	    23	&	1.7353	&	0.4554	&	0.1096	&	0.0012	&	0.2220	&	0.0088  \\
{}	&	     9	&	    27	&	1.7147	&	0.3786	&	0.1079	&	0.0011	&	0.2090	&	0.0033  \\[3mm]
40	&	     7	&	    32	&	1.7028	&	0.3272	&	0.1105	&	0.0012	&	0.2081	&	0.0026  \\
{}	&	     9	&	    32	&	1.6905	&	0.3092	&	0.1079	&	0.0009	&	0.2083	&	0.0031  \\
{}	&	     9	&	    36	&	1.6659	&	0.2506	&	0.1073	&	0.0009	&	0.2061	&	0.0023  \\[3mm]
50	&	     7	&	    42	&	1.6759	&	0.2408	&	0.1094	&	0.0010	&	0.2049	&	0.0018  \\
{}	&	     9	&	    42	&	1.6440	&	0.2089	&	0.1073	&	0.0007	&	0.2049	&	0.0021  \\
{}	&	     9	&	    45	&	1.6284	&	0.1846	&	0.1058	&	0.0007	&	0.2036	&	0.0018  \\
\bottomrule
\end{tabular}
\label{tab:AE5}
\end{table}
\begin{landscape}
\begin{table}[!ht]\scriptsize
\caption{CPs and ALs of 95\% CRI for $\alpha,$ $\theta_1$ and $\theta_2$ based on $5000$ simulations with $\alpha=1.5$, $\theta_1=0.1$,
and $\theta_2=0.2$ for different values of $n$, $\tau_1$ and $\tau_2$ of Type-I censored data.}
\centering
\begin{tabular}{*{21}{c}}
\toprule
\multicolumn{3}{c}{} & \multicolumn{6}{c}{$\alpha$} & \multicolumn{6}{c}{$\theta_1$} & \multicolumn{6}{c}{$\theta_2$} \\
\cmidrule(lr){4-9}\cmidrule(lr){10-15}\cmidrule(lr){16-21}
\multicolumn{3}{c}{} & \multicolumn{2}{c}{Left CRI} & \multicolumn{2}{c}{Symmetric CRI} &  \multicolumn{2}{c}{HPD CRI}
& \multicolumn{2}{c}{Left CRI} & \multicolumn{2}{c}{Symmetric CRI} &  \multicolumn{2}{c}{HPD CRI}
& \multicolumn{2}{c}{Left CRI} & \multicolumn{2}{c}{Symmetric CRI} &  \multicolumn{2}{c}{HPD CRI}\\
\cmidrule(lr){4-5}\cmidrule(lr){6-7}\cmidrule(lr){8-9}\cmidrule(lr){10-11}\cmidrule(lr){12-13}\cmidrule(lr){14-15}\cmidrule(lr){16-17}\cmidrule(lr){18-19}\cmidrule(lr){20-21}  
\multicolumn{1}{c}{$n$} & \multicolumn{1}{c}{$\tau_1$} & \multicolumn{1}{c}{$\tau_2$} & \multicolumn{1}{c}{CP} & \multicolumn{1}{c}{AL} &
\multicolumn{1}{c}{CP} & \multicolumn{1}{c}{AL} & \multicolumn{1}{c}{CP} & \multicolumn{1}{c}{AL} &
\multicolumn{1}{c}{CP} & \multicolumn{1}{c}{AL} & \multicolumn{1}{c}{CP} & \multicolumn{1}{c}{AL} &
\multicolumn{1}{c}{CP} & \multicolumn{1}{c}{AL} & \multicolumn{1}{c}{CP} & \multicolumn{1}{c}{AL} &
\multicolumn{1}{c}{CP} & \multicolumn{1}{c}{AL} & \multicolumn{1}{c}{CP} & \multicolumn{1}{c}{AL}\\
\midrule
20	&	     7	&	    13	&	 95.64	&	3.1864	&	 97.12	&	3.0328	&	 94.28	&	2.7679	&	 97.78	&	0.1947	&	 98.14	&	0.1743	&	 95.78	&	0.1629	&	 94.24	&	0.2726	&	 95.38	&	0.2530	&	 93.22	&	0.2408  \\
{}	&	     9	&	    13	&	 95.44	&	3.1317	&	 96.28	&	2.9208	&	 93.46	&	2.6716	&	 96.80	&	0.1840	&	 97.00	&	0.1561	&	 93.88	&	0.1455	&	 95.08	&	0.3055	&	 96.02	&	0.2982	&	 94.20	&	0.2800  \\
{}	&	     9	&	    15	&	 95.16	&	3.0708	&	 96.30	&	2.8365	&	 93.02	&	2.6166	&	 96.64	&	0.1803	&	 97.22	&	0.1541	&	 93.68	&	0.1445	&	 94.70	&	0.2871	&	 95.92	&	0.2680	&	 93.96	&	0.2547  \\[3mm]
30	&	     7	&	    13	&	 96.40	&	2.7229	&	 97.18	&	2.3466	&	 93.80	&	2.1774	&	 98.04	&	0.1802	&	 97.44	&	0.1466	&	 94.94	&	0.1375	&	 94.02	&	0.2326	&	 95.40	&	0.2043	&	 92.98	&	0.1961  \\
{}	&	     9	&	    13	&	 95.63	&	2.6491	&	 96.20	&	2.2070	&	 92.80	&	2.0526	&	 97.23	&	0.1688	&	 96.57	&	0.1271	&	 94.13	&	0.1187	&	 94.40	&	0.2556	&	 95.70	&	0.2392	&	 93.63	&	0.2268  \\
{}	&	     9	&	    15	&	 95.53	&	2.5628	&	 96.00	&	2.1399	&	 93.47	&	2.0030	&	 96.83	&	0.1670	&	 97.37	&	0.1293	&	 94.47	&	0.1219	&	 94.47	&	0.2461	&	 95.77	&	0.2192	&	 94.13	&	0.2103  \\[3mm]
40	&	     7	&	    13	&	 96.47	&	2.4529	&	 96.37	&	1.9535	&	 93.23	&	1.8261	&	 97.73	&	0.1709	&	 97.20	&	0.1283	&	 94.63	&	0.1207	&	 94.67	&	0.2092	&	 95.53	&	0.1772	&	 94.00	&	0.1705  \\
{}	&	     9	&	    13	&	 95.30	&	2.3766	&	 95.70	&	1.8153	&	 92.60	&	1.6960	&	 97.17	&	0.1595	&	 96.67	&	0.1088	&	 94.50	&	0.1015	&	 93.90	&	0.2285	&	 95.50	&	0.2077	&	 92.93	&	0.1978  \\
{}	&	     9	&	    15	&	 96.10	&	2.3392	&	 96.20	&	1.7972	&	 93.67	&	1.6899	&	 96.87	&	0.1584	&	 96.93	&	0.1123	&	 93.93	&	0.1058	&	 93.57	&	0.2171	&	 95.97	&	0.1867	&	 93.53	&	0.1800  \\[3mm]
50	&	     7	&	    13	&	 96.80	&	2.2919	&	 96.63	&	1.7071	&	 93.57	&	1.5995	&	 97.83	&	0.1650	&	 97.13	&	0.1159	&	 94.80	&	0.1089	&	 94.17	&	0.1911	&	 95.13	&	0.1573	&	 93.40	&	0.1517  \\
{}	&	     9	&	    13	&	 96.37	&	2.2664	&	 95.47	&	1.6042	&	 92.33	&	1.5057	&	 98.13	&	0.1562	&	 97.07	&	0.0979	&	 95.27	&	0.0914	&	 94.30	&	0.2102	&	 95.37	&	0.1871	&	 93.73	&	0.1784  \\
{}	&	     9	&	    15	&	 96.23	&	2.2150	&	 96.23	&	1.5880	&	 93.33	&	1.4986	&	 97.23	&	0.1544	&	 96.77	&	0.1019	&	 93.50	&	0.0962	&	 94.20	&	0.2030	&	 95.33	&	0.1693	&	 93.67	&	0.1636  \\
\bottomrule
\end{tabular}
\label{tab:CRI4}
\end{table}
\end{landscape}

\begin{landscape}
\begin{table}[!ht]\scriptsize
\caption{CPs and ALs of 95\% CRI for $\alpha,$ $\theta_1$ and $\theta_2$ based on $5000$ simulations with $\alpha=1.5$, $\theta_1=0.1$,
and $\theta_2=0.2$ for different values of $n$, $\tau_1$ and $r$ of Type-II censored data.}
\centering
\begin{tabular}{*{21}{c}}
\toprule
\multicolumn{3}{c}{} & \multicolumn{6}{c}{$\alpha$} & \multicolumn{6}{c}{$\theta_1$} & \multicolumn{6}{c}{$\theta_2$} \\
\cmidrule(lr){4-9}\cmidrule(lr){10-15}\cmidrule(lr){16-21}
\multicolumn{3}{c}{} & \multicolumn{2}{c}{Left CRI} & \multicolumn{2}{c}{Symmetric CRI} &  \multicolumn{2}{c}{HPD CRI}
& \multicolumn{2}{c}{Left CRI} & \multicolumn{2}{c}{Symmetric CRI} &  \multicolumn{2}{c}{HPD CRI}
& \multicolumn{2}{c}{Left CRI} & \multicolumn{2}{c}{Symmetric CRI} &  \multicolumn{2}{c}{HPD CRI}\\
\cmidrule(lr){4-5}\cmidrule(lr){6-7}\cmidrule(lr){8-9}\cmidrule(lr){10-11}\cmidrule(lr){12-13}\cmidrule(lr){14-15}\cmidrule(lr){16-17}\cmidrule(lr){18-19}\cmidrule(lr){20-21}  
\multicolumn{1}{c}{$n$} & \multicolumn{1}{c}{$\tau_1$} & \multicolumn{1}{c}{$r$} & \multicolumn{1}{c}{CP} & \multicolumn{1}{c}{AL} &
\multicolumn{1}{c}{CP} & \multicolumn{1}{c}{AL} & \multicolumn{1}{c}{CP} & \multicolumn{1}{c}{AL} &
\multicolumn{1}{c}{CP} & \multicolumn{1}{c}{AL} & \multicolumn{1}{c}{CP} & \multicolumn{1}{c}{AL} &
\multicolumn{1}{c}{CP} & \multicolumn{1}{c}{AL} & \multicolumn{1}{c}{CP} & \multicolumn{1}{c}{AL} &
\multicolumn{1}{c}{CP} & \multicolumn{1}{c}{AL} & \multicolumn{1}{c}{CP} & \multicolumn{1}{c}{AL}\\
\midrule
20	&	     7	&	    15	&	 96.30	&	3.3992	&	 96.98	&	3.2367	&	 94.28	&	2.9648	&	 97.80	&	0.2018	&	 97.64	&	0.1819	&	 95.74	&	0.1700	&	 95.24	&	0.3174	&	 94.98	&	0.3037	&	 93.94	&	0.2866  \\
{}	&	     9	&	    15	&	 95.76	&	3.2467	&	 96.10	&	3.0317	&	 93.20	&	2.7877	&	 97.40	&	0.1878	&	 97.26	&	0.1617	&	 95.46	&	0.1513	&	 95.66	&	0.3764	&	 96.16	&	0.3808	&	 95.40	&	0.3513  \\
{}	&	     9	&	    17	&	 95.50	&	3.0823	&	 96.16	&	2.8396	&	 93.02	&	2.6304	&	 97.48	&	0.1838	&	 97.06	&	0.1584	&	 94.32	&	0.1496	&	 95.40	&	0.3124	&	 95.64	&	0.2924	&	 94.96	&	0.2782  \\[3mm]
30	&	     7	&	    23	&	 96.28	&	2.7953	&	 96.82	&	2.4164	&	 93.64	&	2.2468	&	 98.02	&	0.1845	&	 97.14	&	0.1509	&	 95.16	&	0.1419	&	 94.80	&	0.2508	&	 95.32	&	0.2245	&	 94.02	&	0.2154  \\
{}	&	     9	&	    23	&	 96.38	&	2.6635	&	 96.06	&	2.2360	&	 93.52	&	2.0851	&	 97.44	&	0.1715	&	 96.58	&	0.1321	&	 95.02	&	0.1241	&	 95.42	&	0.2862	&	 96.06	&	0.2708	&	 95.28	&	0.2558  \\
{}	&	     9	&	    27	&	 96.08	&	2.5628	&	 96.16	&	2.1224	&	 93.24	&	2.0007	&	 96.88	&	0.1671	&	 96.58	&	0.1307	&	 93.10	&	0.1243	&	 94.78	&	0.2478	&	 95.66	&	0.2105	&	 94.46	&	0.2035  \\[3mm]
40	&	     7	&	    32	&	 96.54	&	2.4596	&	 96.36	&	1.9713	&	 93.48	&	1.8483	&	 97.62	&	0.1721	&	 97.06	&	0.1319	&	 94.98	&	0.1244	&	 94.46	&	0.2163	&	 94.88	&	0.1826	&	 93.76	&	0.1764  \\
{}	&	     9	&	    32	&	 96.46	&	2.4068	&	 95.78	&	1.8552	&	 92.76	&	1.7464	&	 97.06	&	0.1610	&	 96.60	&	0.1148	&	 93.88	&	0.1083	&	 94.44	&	0.2333	&	 95.70	&	0.2050	&	 94.48	&	0.1974  \\
{}	&	     9	&	    36	&	 96.26	&	2.3261	&	 96.50	&	1.7818	&	 93.58	&	1.6901	&	 97.28	&	0.1591	&	 96.44	&	0.1157	&	 93.58	&	0.1101	&	 94.42	&	0.2233	&	 95.16	&	0.1810	&	 94.40	&	0.1758  \\[3mm]
50	&	     7	&	    42	&	 96.46	&	2.2933	&	 96.38	&	1.7292	&	 93.18	&	1.6310	&	 97.42	&	0.1650	&	 96.80	&	0.1208	&	 93.92	&	0.1144	&	 94.32	&	0.1965	&	 95.08	&	0.1553	&	 93.80	&	0.1508  \\
{}	&	     9	&	    42	&	 96.22	&	2.2202	&	 95.74	&	1.6018	&	 92.68	&	1.5161	&	 97.14	&	0.1549	&	 96.30	&	0.1045	&	 93.24	&	0.0990	&	 94.52	&	0.2101	&	 95.48	&	0.1735	&	 94.74	&	0.1684  \\
{}	&	     9	&	    45	&	 96.48	&	2.1732	&	 96.44	&	1.5640	&	 93.70	&	1.4855	&	 97.32	&	0.1525	&	 96.48	&	0.1046	&	 94.00	&	0.0995	&	 93.66	&	0.2065	&	 95.18	&	0.1610	&	 94.38	&	0.1569  \\
\bottomrule
\end{tabular}
\label{tab:CRI5}
\end{table}
\end{landscape}

\end{document}